\documentclass[journal]{IEEEtran}
\IEEEoverridecommandlockouts
\usepackage{amsmath,amssymb,amsfonts}
\usepackage{algorithmic}
\usepackage{graphicx}
\usepackage{textcomp}
\usepackage{xcolor}
\usepackage{soul,enumitem}
\usepackage{hyperref}
\usepackage[comma,sort&compress,numbers]{natbib}
\usepackage{booktabs}
\usepackage{gensymb}
	\usepackage{multirow}
	\usepackage{sidecap}
	\usepackage{graphicx}
\usepackage[font=small]{caption}

\definecolor{myred}{rgb}{0.6350, 0.0780, 0.1840}
\definecolor{myyellow}{rgb}{0.9290, 0.6940, 0.1250}
\definecolor{mygreen}{rgb}{0.4660, 0.6740, 0.1880}

\usepackage[linesnumbered,ruled]{algorithm2e}

\def\BibTeX{{\rm B\kern-.05em{\sc i\kern-.025em b}\kern-.08em
    T\kern-.1667em\lower.7ex\hbox{E}\kern-.125emX}}
\begin{document}

\title{BVI-UGC: A Video Quality Database for User-Generated Content Transcoding}
\author{Zihao Qi, ~\IEEEmembership{Student Member,~IEEE,}, Chen Feng,~\IEEEmembership{Student Member,~IEEE,}, Fan Zhang,~\IEEEmembership{Member,~IEEE,}, Xiaozhong Xu,~\IEEEmembership{Senior Member,~IEEE,}, Shan Liu,~\IEEEmembership{Fellow,~IEEE,} and David R. Bull,~\IEEEmembership{Fellow,~IEEE}
\thanks{Zihao Qi, Chen Feng, Fan Zhang and David Bull are with the Visual Information Lab, University of Bristol, Bristol BS1 5DD, U.K. (e-mail: \{zihao.qi, chen.feng, fan.zhang, dave.bull@bristol.ac.uk). Xiaozhong Xu and Shan Liu are with Tencent America, Palo Alto, USA. (e-mail: \{xiaozhongxu, shanl\}@tencent.com}

\thanks{The authors acknowledge the funding from Tencent (US), University of Bristol, and the UKRI MyWorld Strength in Places Programme (SIPF00006/1).}

\thanks{This work involved collecting data from human participants. The relevant experiments have been approved by the Faculty of Engineering Research Ethics Committee of the University of Bristol (Ref 12352).}}
\maketitle

\begin{abstract} 
In recent years, user-generated content (UGC) has become one of the major video types consumed  via streaming networks. Numerous research contributions have focused on assessing its visual quality through subjective tests and objective modeling. In most cases, objective assessments are based on a no-reference scenario, where the corresponding reference content is assumed not to be available. However, full-reference video quality assessment is also important for UGC in the delivery pipeline, particularly associated with the video transcoding process. In this context, we present a new UGC video quality database, BVI-UGC, for user-generated content transcoding, which contains 60 (non-pristine) reference videos and 1,080 test sequences. In this work, we simulated the creation of non-pristine reference sequences (with a wide range of compression distortions), typical of content uploaded to UGC platforms for transcoding. A comprehensive crowdsourced subjective study was then conducted involving more than 3,500 human participants. Based on this collected subjective data, we benchmarked the performance of 10 full-reference and 11 no-reference quality metrics. Our results demonstrate the poor performance (SROCC values are lower than 0.6) of these metrics in predicting the perceptual quality of UGC in two different scenarios (with or without a reference).
To facilitate future research in this area, we have made BVI-UGC publicly available at \url{https://zihaoq1.github.io/BVI-UGC/}
\end{abstract}

\begin{IEEEkeywords}
Video quality assessment, UGC, video transcoding, BVI-UGC, subjective study, crowdsourcing
\end{IEEEkeywords}

\section{Introduction}\label{sec:intro}

With advances in mobile devices and communication network technologies, coupled with the explosion of social media and streaming platforms, user-generated content (UGC) now represents more than 35\% of downstream volume over fixed networks~\cite{sandvine2024}, streamed by service providers such as YouTube, Facebook, TikTok, and Tencent. UGC also significantly influences upstream traffic, driven by the popularity of short-form videos~\cite{sandvine2024}. These highlight the importance of video compression in UGC streaming, which plays a critical role in managing the trade-offs between video quality and required bandwidth.

\begin{figure}[t]
  \centering
  \includegraphics[width=\linewidth]{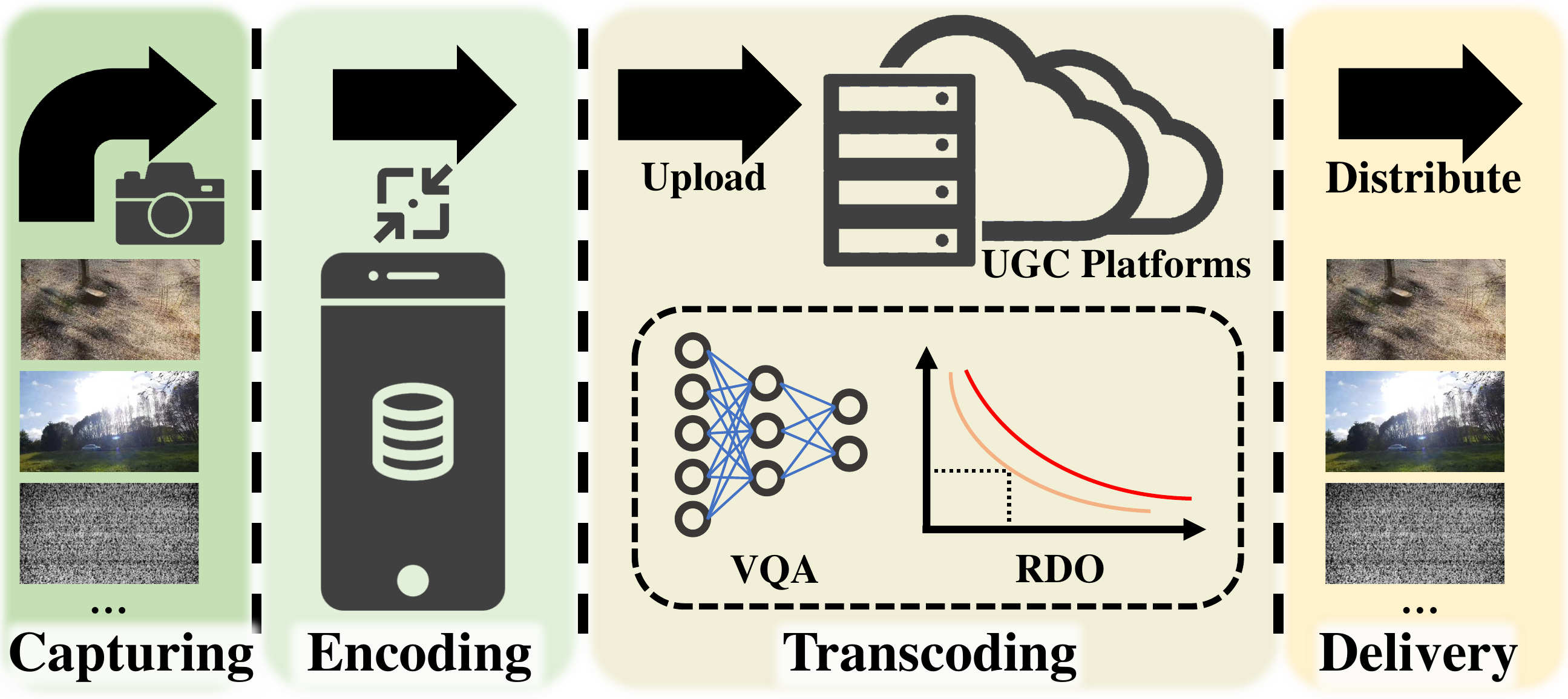} % 
  \caption{Illustration of the UGC video delivery pipeline. Source videos captured by users may contain various distortions due to poor quality equipment, unskilled cinematography and lossy compression. Captured videos are uploaded to UGC platforms where they are transcoded and streamed to video consumers. During the transcoding process, a perceptually accurate VQA metric is a key component in rate-distortion optimization.}
\label{fig:transcoding}
\end{figure}

Compared to professionally-generated content (PGC), UGC has unique characteristics due to the specific production and delivery pipeline employed (~\autoref{fig:transcoding}). UGC videos are typically captured using commercial and mobile devices by amateur users. In most cases, the original uncompressed sources of these videos are not stored during acquisition, as they are directly compressed using a fast video codec embedded in the device (e.g., x264~\cite{x264repos}). The compressed videos are then uploaded to a selected UGC platform for streaming, where a further layer of compression (i.e., transcoding) is performed. Due to the nature of UGC, its transcoding is often based on references, which themselves contain significant source and compression artifacts. This differs from the PGC production pipeline, where high-quality original content is always available during encoding.  

In current transcoding pipelines used by many UGC platforms, the encoding operation is similar to that used for PGC, where the objective is to minimize the distortion (or the visual quality degradation) between the reference and reconstructed videos. As mentioned above, in UGC transcoding, the input reference content may contain various source and compression artifacts, which reduces the effectiveness of the rate-distortion (or rate-quality) optimization due to the inaccurate quality (distortion) prediction in these cases \cite{wang2019youtube}.  

In recent years, there have been many research contributions addressing the issues associated with UGC-based video quality assessment. However, most solutions~\cite{tu2021ugc,wang2019youtube,konvid1k,hosu2017konstanz,konvid150k,hahn2021,sinno2018large,wu2022fast,wu2023neighbourhood} have been developed for no-reference scenarios, which are typically employed to assess visual quality at the user end. Although there are existing studies that address the subjective quality assessment of transcoded UGC, most of these only consider a limited range of reference quality levels~\cite{li2020ugc,yu2021predicting,wang2021challenge}, which do not reflect real-world scenarios where the quality of reference content varies considerably.

In this context, to facilitate advances in video quality assessment for UGC, we have developed a large-scale UGC database, BVI-UGC, focused on transcoding applications. We collected 60 high-quality source sequences, covering 15 major UGC categories. These videos were then compressed using the x264 codec~\cite{x264repos} with three quantization levels that simulate the compression operation during content capture. The ingested content (i.e., non-pristine reference used in the trascoding stage) was further transcoded using three video codecs (x264, x265~\cite{x265repos} and libaom~\cite{aom}) with two resolution re-sampling levels and three quantization parameters, resulting in a total number of 1,080 distorted test sequences. Moreover, an extensive crowdsourcing-based subjective experiment was performed to collect quality scores from more than 3,500 participants for both test sequences and non-pristine references. The ground-truth data has been further employed to benchmark 21 existing full-reference (FR) and no-reference (NR) objective quality models. The experimental results highlight the poor and inconsistent performance of existing metrics and confirm the urgent requirement for more accurate quality assessment methods. The primary contributions of this work are summarized below.
\begin{enumerate}[leftmargin=*]
    \item BVI-UGC is the \textbf{first public UGC video quality databases containing reference clips compressed with various quantization levels}, which simulates the UGC delivery pipeline. All the source, non-pristine references and test sequences are made available for public evaluation.

    \item All test and reference sequences are labeled with ground truth subjective quality scores through a \textbf{reliable, large-scale crowdsourcing-based psychophysical experiment} using Amazon Mechanical Turk~\cite{turk2023amazon} platform. We also open-sourced the web application used for this subjective test.
    
    \item We demonstrate how to exploit this database by \textbf{benchmarking FR and NR VQA methods in the context of UGC transcoding}. We have also employed this database to evaluate the performance of NR quality metrics in terms of directly measuring the perceptual quality of test sequences without references. Based on a comprehensive experiment involving 21 full-/no-reference metrics, we highlight the challenging nature (no existing quality metrics achieve a SROCC value higher than 0.6 on this dataset) of UGC-based video quality assessment in the context of transcoding applications.
\end{enumerate}

The remainder of this paper is organized as follows. \autoref{sec:related} provides a brief summary of related work in the research areas of UGC video quality assessment and existing UGC databases. \autoref{sec:pro} describes the development of the database, including source video collection and capture, and reference/test sequence generation. \autoref{sec:exp} presents the detailed methodology and configuration employed in the subjective test, and performs an analysis of the collected subjective quality scores. \autoref{sec:metrics} summarizes the results of the benchmark experiment for 21 video quality metrics. Finally, \autoref{sec:conclusion} provides a conclusion of the paper and outlines future work.

\begin{table*}[htb]  
\centering

\caption{Features of notable user-generated content video databases with transcoded sequences. Here ``{L}'' stands for `Landscape' layout, while ``{P}'' stands for `Portrait'. \textsuperscript{\dag}BVI-UGC contains source videos collected from YouTube-UGC database and captured by lab participants using various devices.}
\resizebox{0.98\textwidth}{!}{\begin{tabular}{l|c|c|c|c|c||c} 
\toprule
 & YT-UGC VP9 & LIVE-WILD & UGC-VIDEO & ICME 2021 & TaoLive & \textbf{BVI-UGC}  \\
\midrule
source seq. & n/a & n/a & n/a & n/a & n/a& 60\\
reference seq. & 169 & 55 & 50 & 1000 & 418 & 60\\
transcoded seq. & 567 & 220 & 500 & 7000 & 3,344 & 1080\\
ref. quality & {low} & {medium} & {medium} & {high} & {high} & {3 levels} \\
codecs & VP9 & x264 & x264 and x265 & x264 & x265 & x264,x265 and libaom\\
content source & YouTube & Mobile captured & Mobile captured & Mobile captured & TaoLive & Mixed\textsuperscript{\dag} \\
resolution & 720p,1080p & 360p,540p,720p,1080p & 720p & 720p & 720p,1080p & 540p,1080p\\
layouts & L & L & P & L \& P & mostly P & L \& P \\
frame rate & 30fps & 24-30fps & 24-30fps & 30fps & 20-30fps & 24-60fps\\
bit depth & 8 & 8 & 8 & 8 & 8 & 8\\
duration & 20s & 10s & 10s & 5s & 8s & 5s \\
rating scale & Continuous 1-5 & 0-100 & Discrete 1-5 & Discrete 1-5 & Discrete 1-5 & 0-100 \\
subject number & n/a & 40 & 28 & n/a & 44 & 3,500+\\
ratings Avg. & n/a & 20 & 28 & $>$50 & 44 & 160\\
\bottomrule  
\end{tabular} } 
\label{tab_dataset}
\end{table*}

\section{Related Works}\label{sec:related}

In this section, we first review previous work addressing full-reference and no-reference video quality assessment (VQA), in particular those developed specifically for UGC. We then summarize existing UGC video quality databases, and highlight the urgent need to develop a more diverse video quality database for UGC transcoding.

\subsection{Video Quality Assessment for UGC}

Although subjective tests provide a gold standard for estimating the perceptual quality of video content, they are not widely employed in practical applications due to their time-consuming, non-real-time and expensive nature~\cite{bull2021intelligent}. Instead, objective video quality assessment methods are frequently used in algorithm benchmarking and optimization. These methods can be classified according to the availability of reference content, into two major categories\footnote{There is another class of objective VQA method, denoted reduced-reference (RR), used when only partial information from the reference is available.}: full-reference (FR) models that provide a quality prediction based on the comparison between a processed sequence and the reference counterpart, and no-reference (NR) models, which directly assess the quality of a sequence without considering any information from its reference content.

\subsubsection{Full-Reference VQA} 

FR VQA models typically measure the difference between a test video sequence and its corresponding reference. Simple models such as PSNR are widely employed across many image and video processing applications, serving as a benchmark metric for algorithm comparison and in loss functions for model optimization (e.g., rate-distortion optimization in compression). To further improve their correlation performance with perceptual quality, researchers have developed perceptually-inspired quality metrics that exploit various characteristics of the human vision system (HVS) such as texture masking \cite{ferwerda1997model}, just noticeable difference \cite{yang2005just} and contrast sensitivity functions \cite{ginsburg2003contrast}. Notable examples include SSIM and its variants~\cite{wang2004image,wang2003multiscale,moorthy2009motion,moorthy2010efficient,zeng20123d,rehman2015display}, ADM~\cite{li2011image}, VIF~\cite{sheikh2006image}, MOVIE~\cite{seshadrinathan2009motion}, MAD~\cite{vu2011spatiotemporal} and PVM~\cite{zhang2015perception}. Moreover, some of these perceptual models have been combined together with various video features through linear regression (e.g. SVM~\cite{cortes1995support}) to achieve even higher correlation performance, with one of the most successful examples being VMAF~\cite{zhang2021enhancing}.

More recently, researchers have focused on deep learning-based solutions for quality assessment. Important contributions in this category include DeepVQA~\cite{kim2018deep}, LPIPS~\cite{zhang2018unreasonable}, C3DVQA~\cite{xu2020c3dvqa}, and CUGCVQA~\cite{li2021full}. Although these methods show promise when compared to conventional and regression-based methods, in most cases they demand an intra-database cross-validation due to their limited model generalization ability. To address this issue, more recently, unsupervised or weakly-supervised learning strategies have been developed for deep VQA models \cite{feng2024rankdvqa,madhusudana2022image,peng2024rmt}, which do not require training data with ground-truth subjective scores.

When employed in the context of UGC transcoding, where reference content is often non-pristine, containing various visible artifacts, the aforementioned FR VQA methods do not offer satisfactory correlation performance with ground-truth subjective data~\cite{wang2019youtube,qi2023full}. This can lead to inconsistent quality prediction when used for algorithm evaluation and poor rate quality optimization performance if employed in the transcoding loop.

\subsubsection{No-reference VQA} 

NR quality metrics are employed when the reference content is not available during quality assessment. In video delivery, this is typically applied at the decoder (user end) to measure the quality of user experience. Numerous NR quality assessment methods have been developed for estimating compression distortion~\cite{zhu2014no,caviedes2017no}, transmission errors~\cite{zhang2012additive,wang2015no} and specific artifacts~\cite{wang2016perceptual,ghadiyaram2017no}. 

Similar to FR VQA, recent NR quality metrics have exploited deep neural networks to enhance quality assessment. These have demonstrated improved performance over conventional NR VQA methods based on classical signal processing theories. These learning-based methods can be classified as supervised or weakly-/un-supervised approaches. The former directly performs model learning based on ground-truth subjective scores collected from human participants in large scale experiments, with notable examples including V-MEON~\cite{liu2018end}, ChipQA~\cite{ebenezer2021chipqa}, Compressed VQA~\cite{li2021full} SimpleVQA~\cite{sun2022deep}, FastVQA~\cite{wu2022fast} and FasterVQA~\cite{wu2023neighbourhood}. Un-supervised or weakly-supervised methods typically convert the main task (directly predicting quality scores) to an auxiliary mission based on techniques such as contrastive learning or ranking learning. This supports the generation of more diverse training content without performing expensive subjective tests. Important work in this class includes NR-RankDVQA~\cite{feng2024rankdvqa},  CONTRIQUE~\cite{madhusudana2022image} and CONVIQT~\cite{madhusudana2023conviqt}.

Many NR quality metrics focus on assessing the quality of the UGC content, such as~\cite{li2021full,tu2021ugc,madhusudana2022image,madhusudana2023conviqt,sun2022deep}. However, their performance has not been fully investigated in the context of UGC transcoding due to the limited availability of benchmark databases of this type. 

\subsection{UGC Video Quality Databases} 
\label{subsec:databases}

To evaluate the performance of objective quality metrics, subjective video quality databases that contain various distorted (and reference, if for full-reference scenarios) sequences with different visual artifacts, are used. Human participants are employed to view these sequences through psychophysical experiments based on certain test methodologies and procedures~\cite{recommendation2022910}. Different correlation coefficients between the quality indices generated by the objective quality models for these test sequences and their corresponding ground-truth subjective scores, collected in the subjective experiment, can then be used to measure and compare the performance of these objective quality models. 

Early work on subjective video databases typically focused on PGC videos, with notable examples such as VQEG FR Phase I, VQEG HD~\cite{vqeg_web}, LIVE-VQA~\cite{seshadrinathan2010study}, IVP~\cite{zhang2011ivp} and BVI-HD~\cite{zhang2018bvi}, which primarily investigate the impact of video compression and transmission. There are also contributions that study the influence of specific video formats or artifacts, including BVI-HFR (frame rate)~\cite{mackin2015study}, LIVE-YT-HFR (frame rate)~\cite{madhusudana2021subjective,madhusudana2020capturing}, BVI-SR (spatial resolution)~\cite{mackin2018study}, BVI-BD (bit depth)~\cite{mackin2021subjective}, BAND-2k (banding artifacts)~\cite{chen2023fs,chen2024band}, etc.

\begin{figure*}[!t]
\footnotesize
\begin{minipage}[b]{0.557\linewidth}
\begin{minipage}[b]{0.24\linewidth}
  \centering
  \centerline{\includegraphics[width=\textwidth]{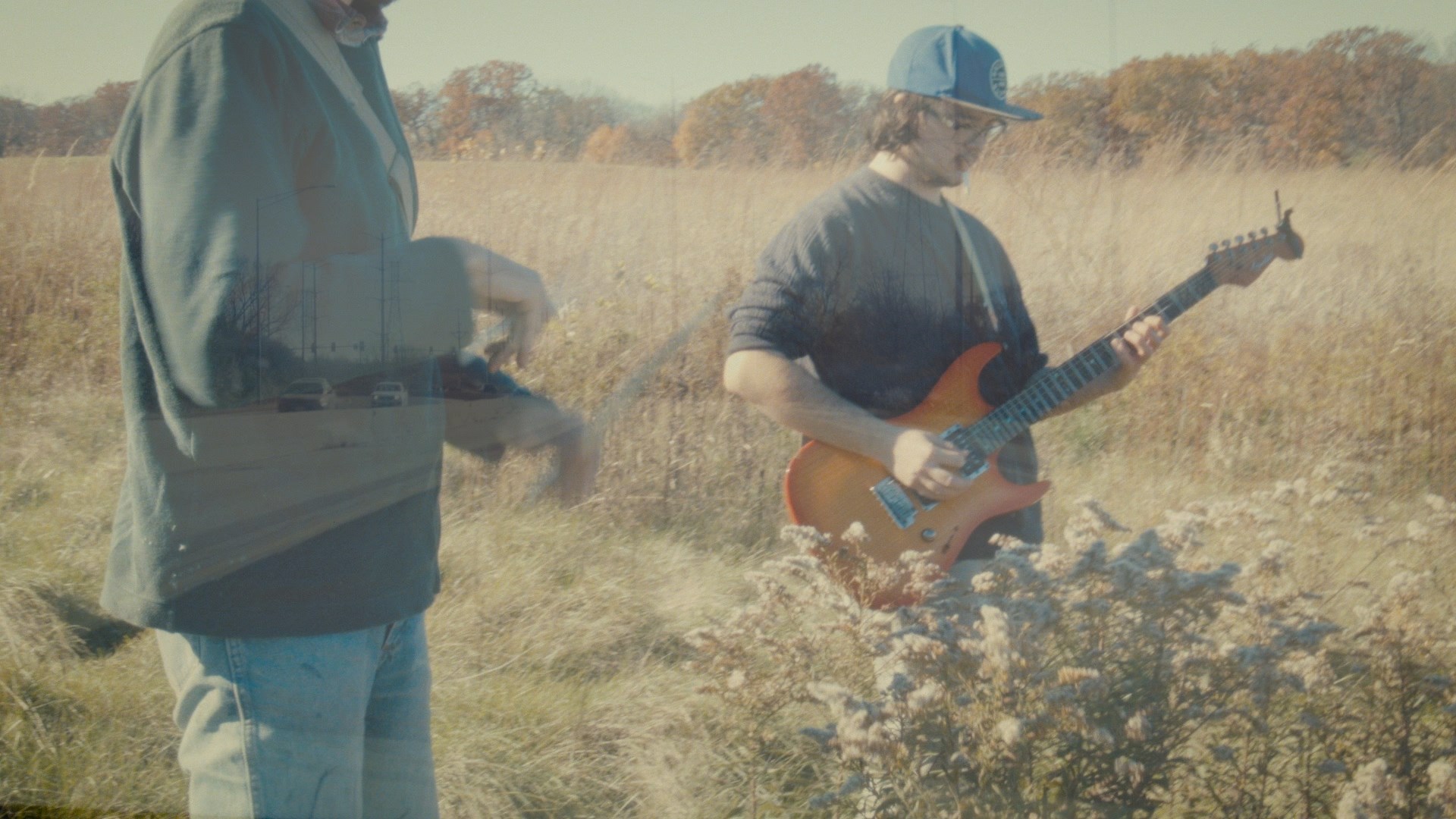}}
\vspace{-3pt}(1) Lyric Video
\end{minipage}
\begin{minipage}[b]{0.24\linewidth}
  \centering
  \centerline{\includegraphics[width=\textwidth]{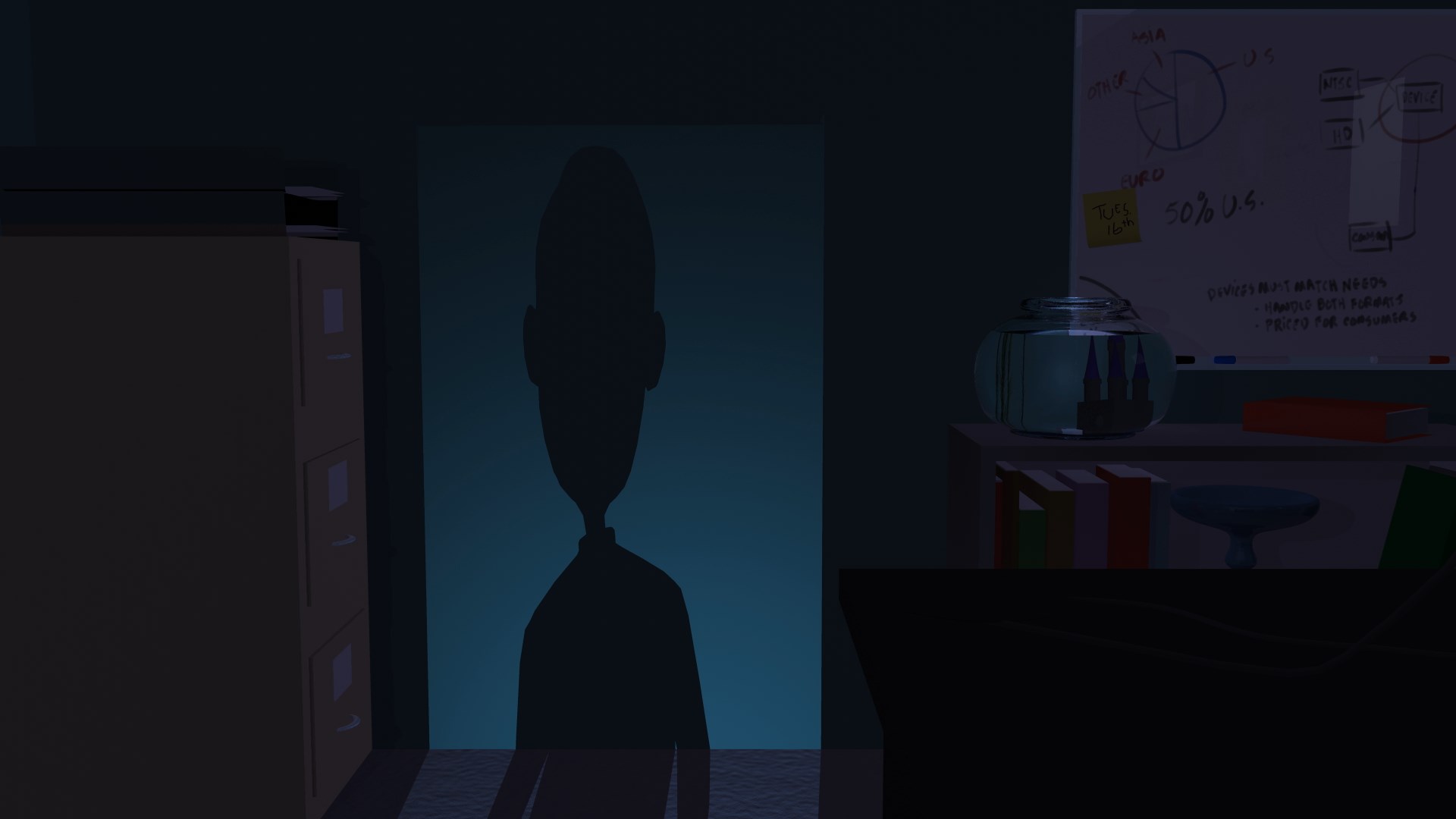}}
\vspace{-3pt}(2) Animation
\end{minipage}
\begin{minipage}[b]{0.24\linewidth}
  \centering
  \centerline{\includegraphics[width=\textwidth]{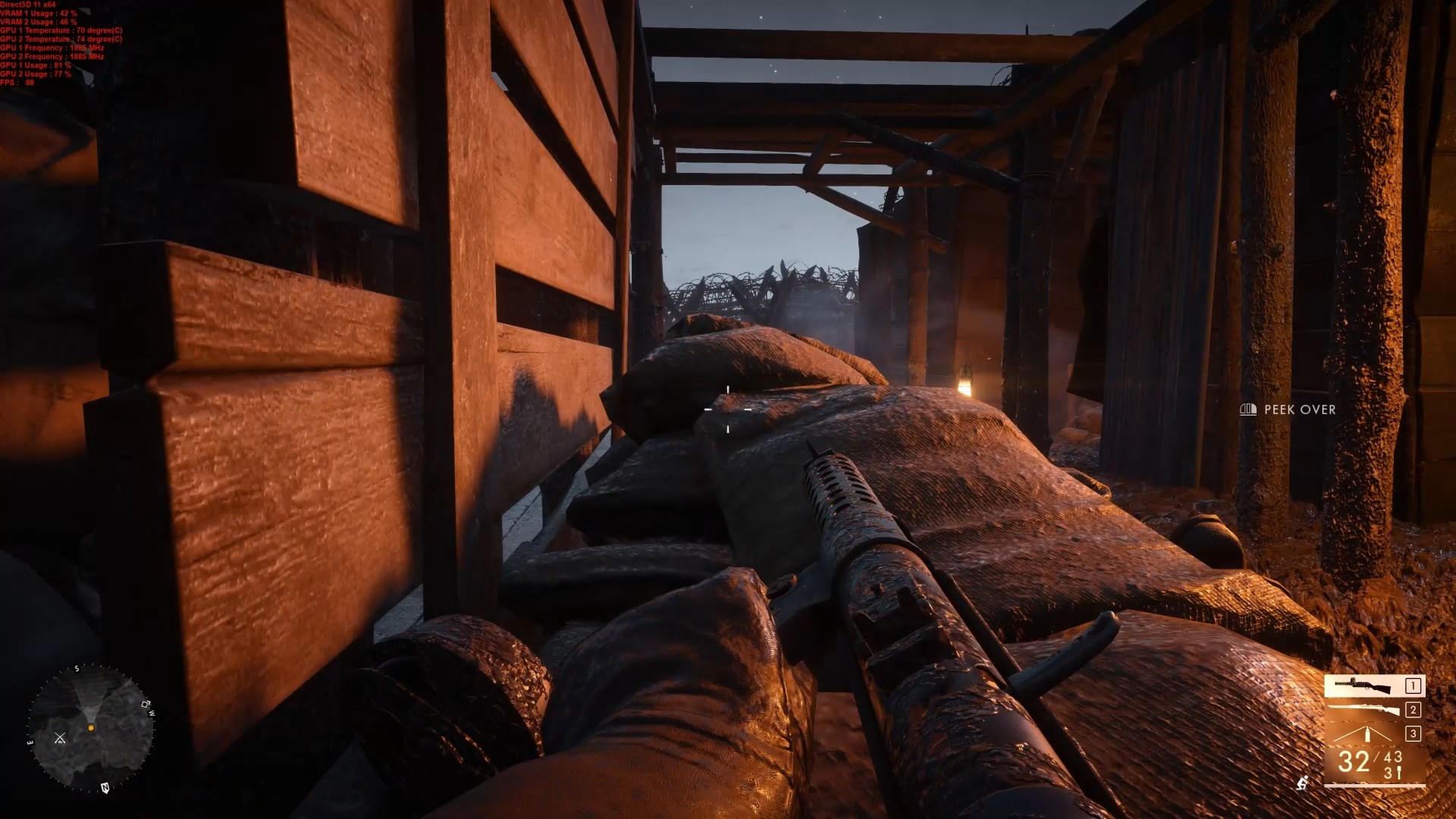}}
\vspace{-3pt}(3) Game
\end{minipage}
\begin{minipage}[b]{0.24\linewidth}
  \centering
  \centerline{\includegraphics[width=\textwidth]{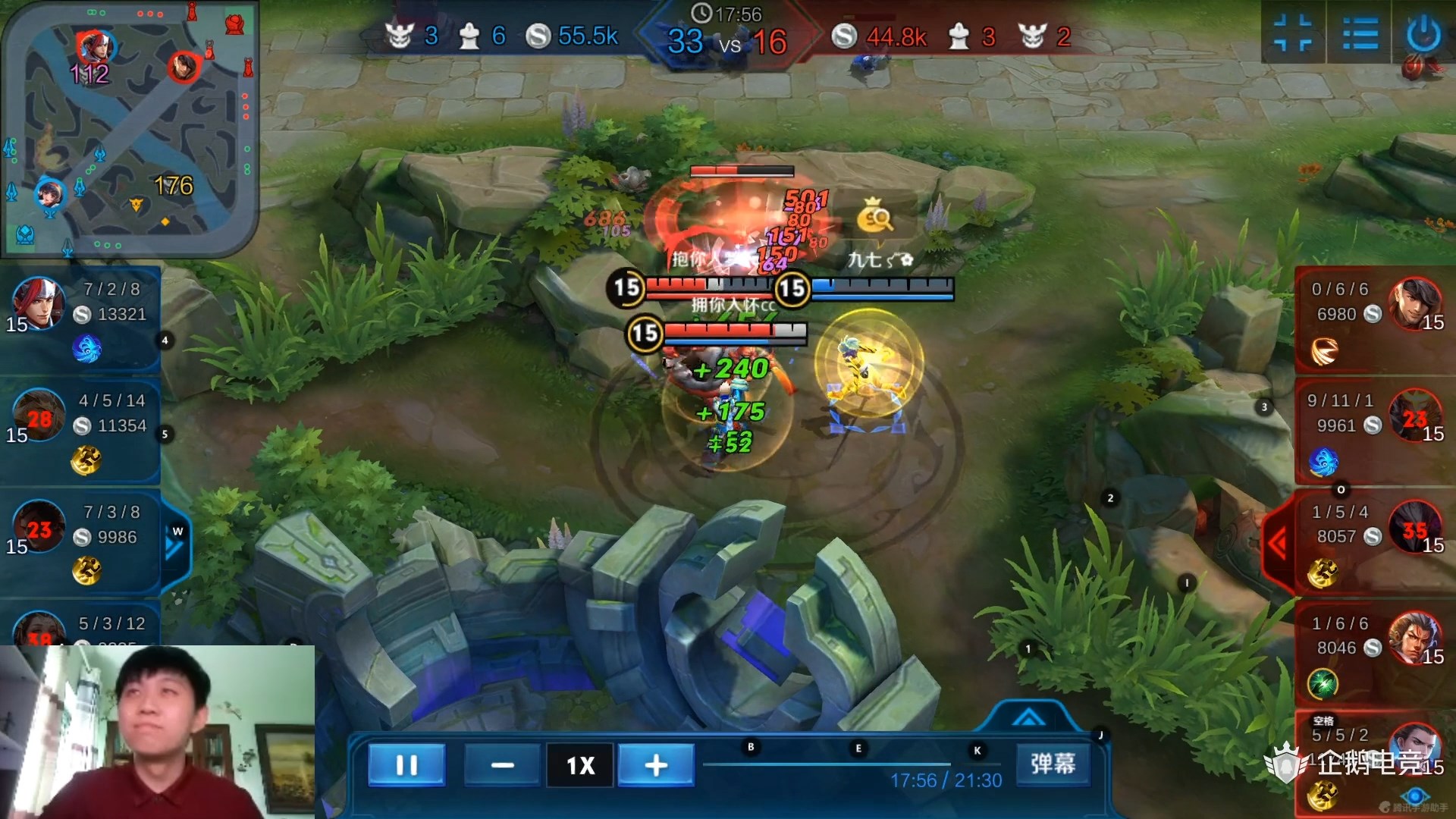}}
\vspace{-3pt}(4) Streamer
\end{minipage}

\begin{minipage}[b]{0.24\linewidth}
  \centering
  \centerline{\includegraphics[width=\textwidth]{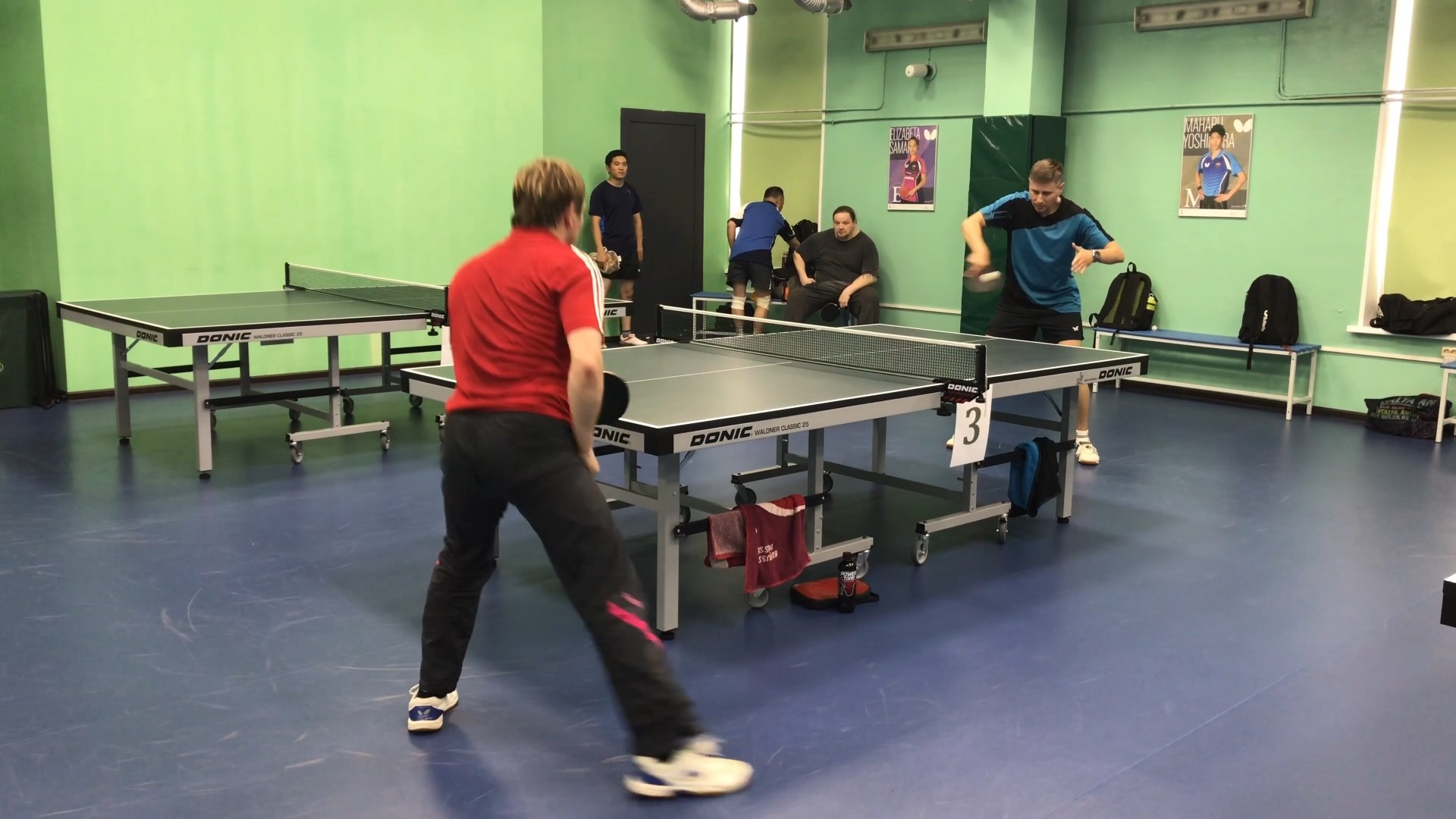}}
\vspace{-3pt}(5) Sports
\end{minipage}
\begin{minipage}[b]{0.24\linewidth}
  \centering
  \centerline{\includegraphics[width=\textwidth]{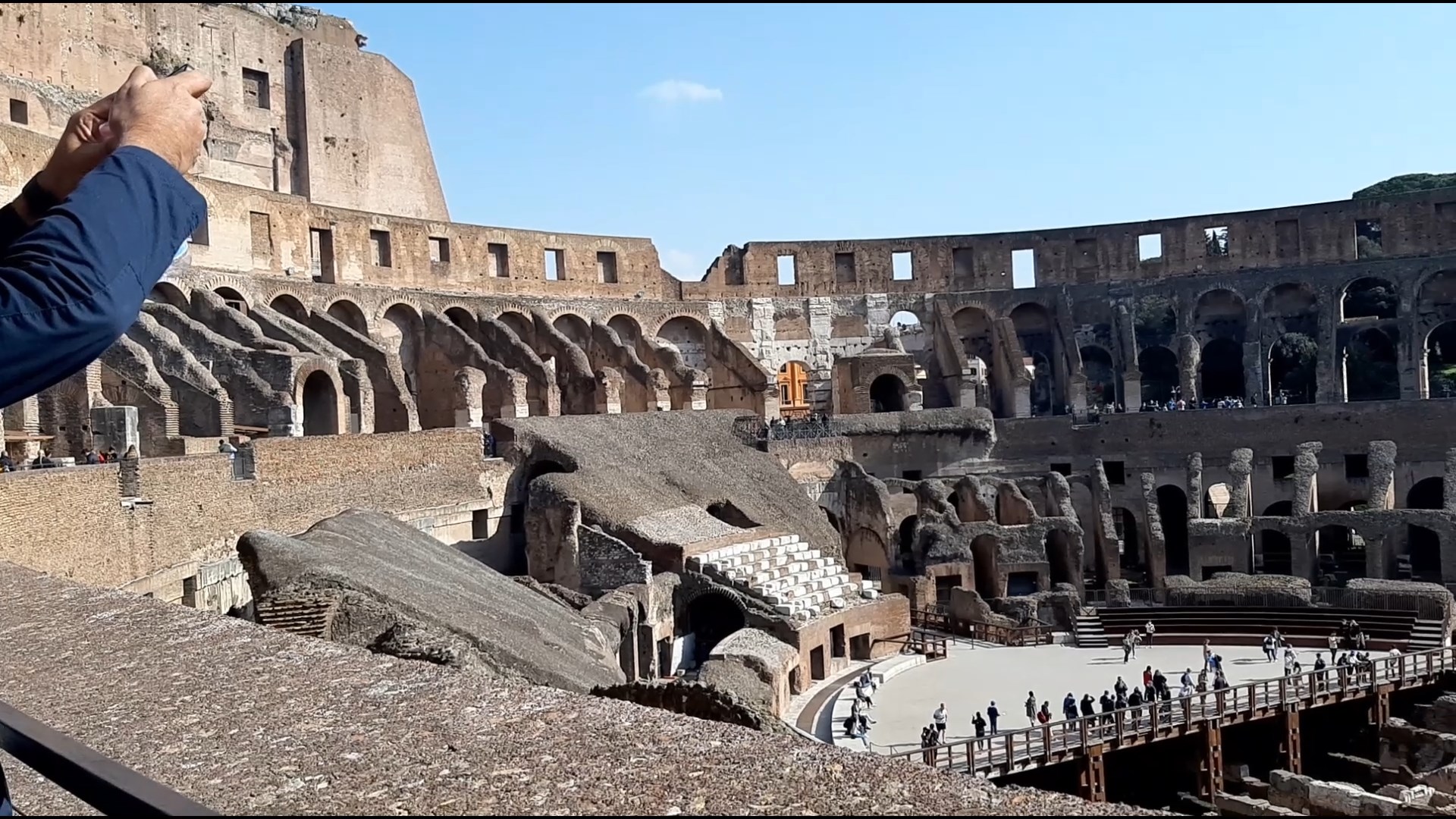}}
\vspace{-3pt}(6) Travel
\end{minipage}
\begin{minipage}[b]{0.24\linewidth}
  \centering
  \centerline{\includegraphics[width=\textwidth]{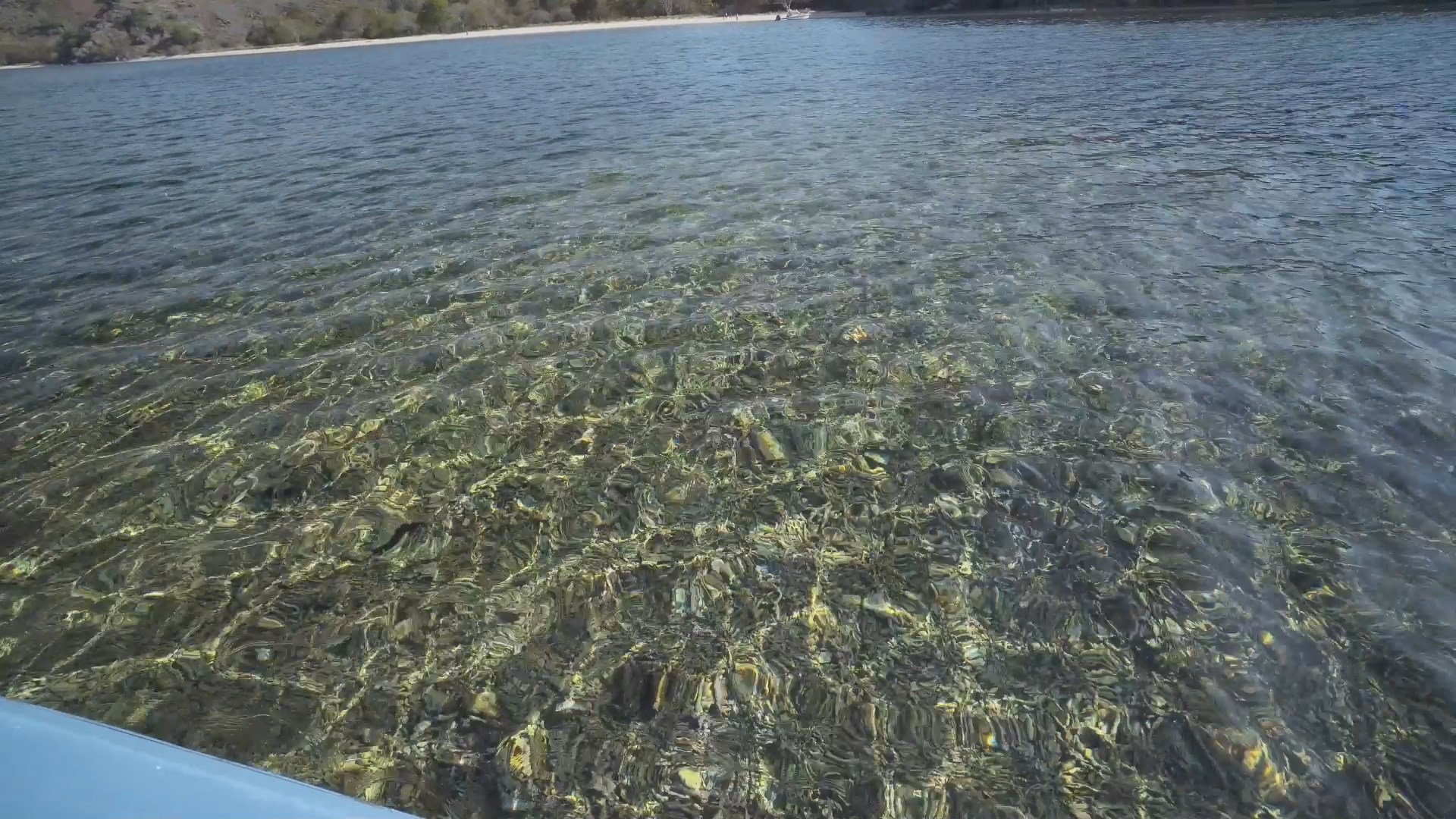}}
\vspace{-3pt}(7) Lifestyle
\end{minipage}
\begin{minipage}[b]{0.24\linewidth}
  \centering
  \centerline{\includegraphics[width=\textwidth]{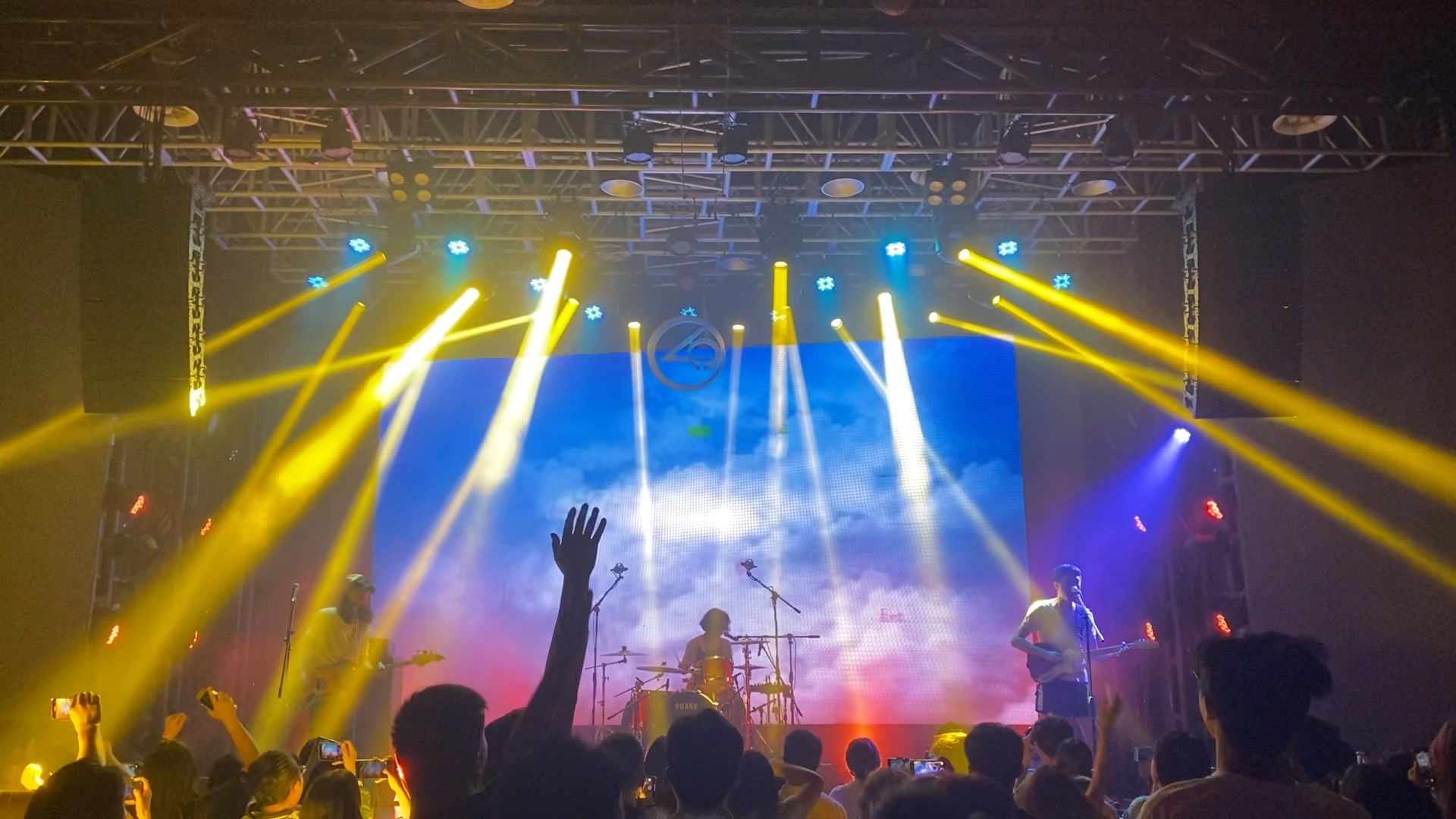}}
\vspace{-3pt}(8) Performance
\end{minipage}

\begin{minipage}[b]{0.24\linewidth}
  \centering
  \centerline{\includegraphics[width=\textwidth]{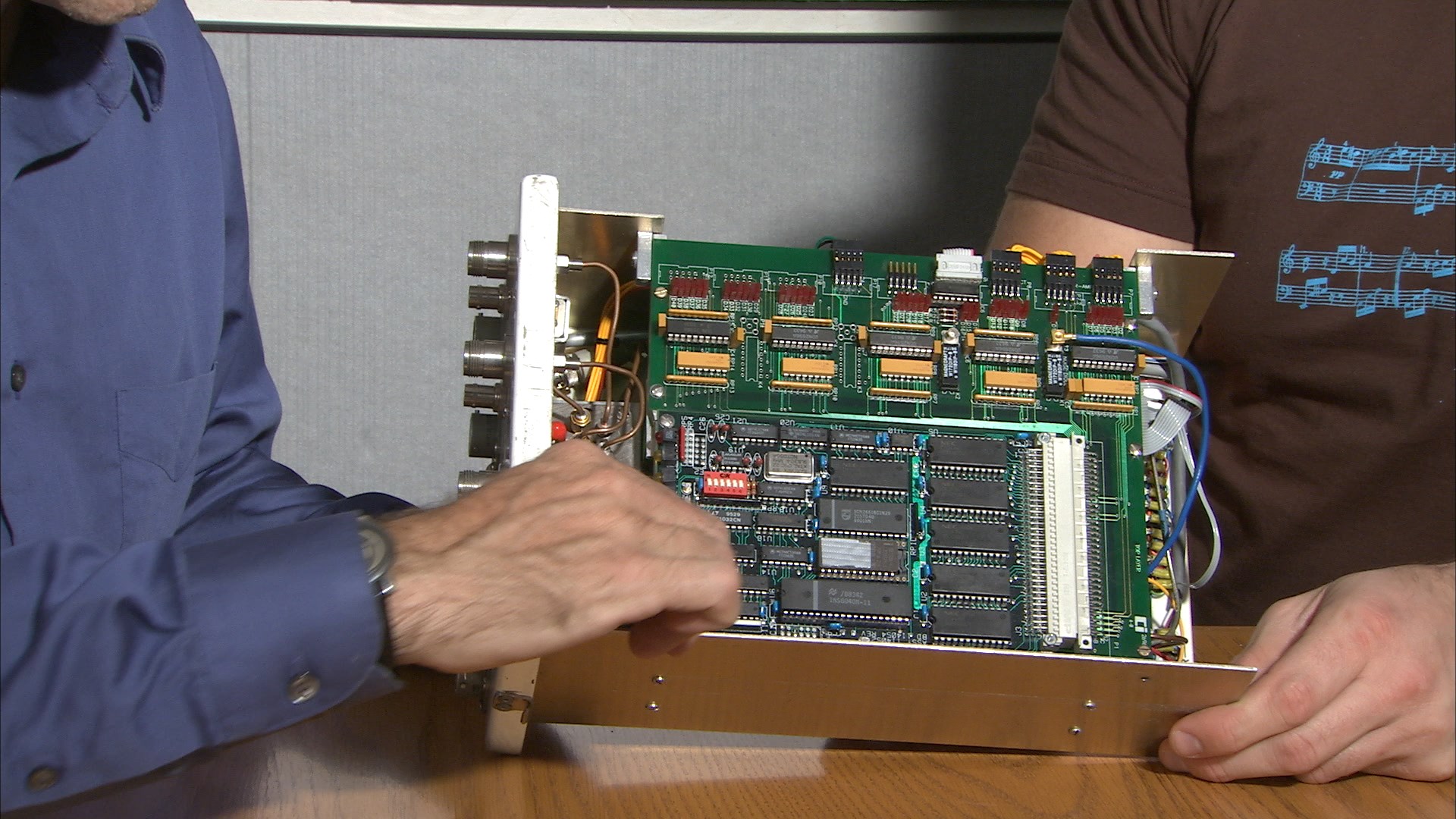}}
\vspace{-3pt}(9) Lecture
\end{minipage}
\begin{minipage}[b]{0.24\linewidth}
  \centering
  \centerline{\includegraphics[width=\textwidth]{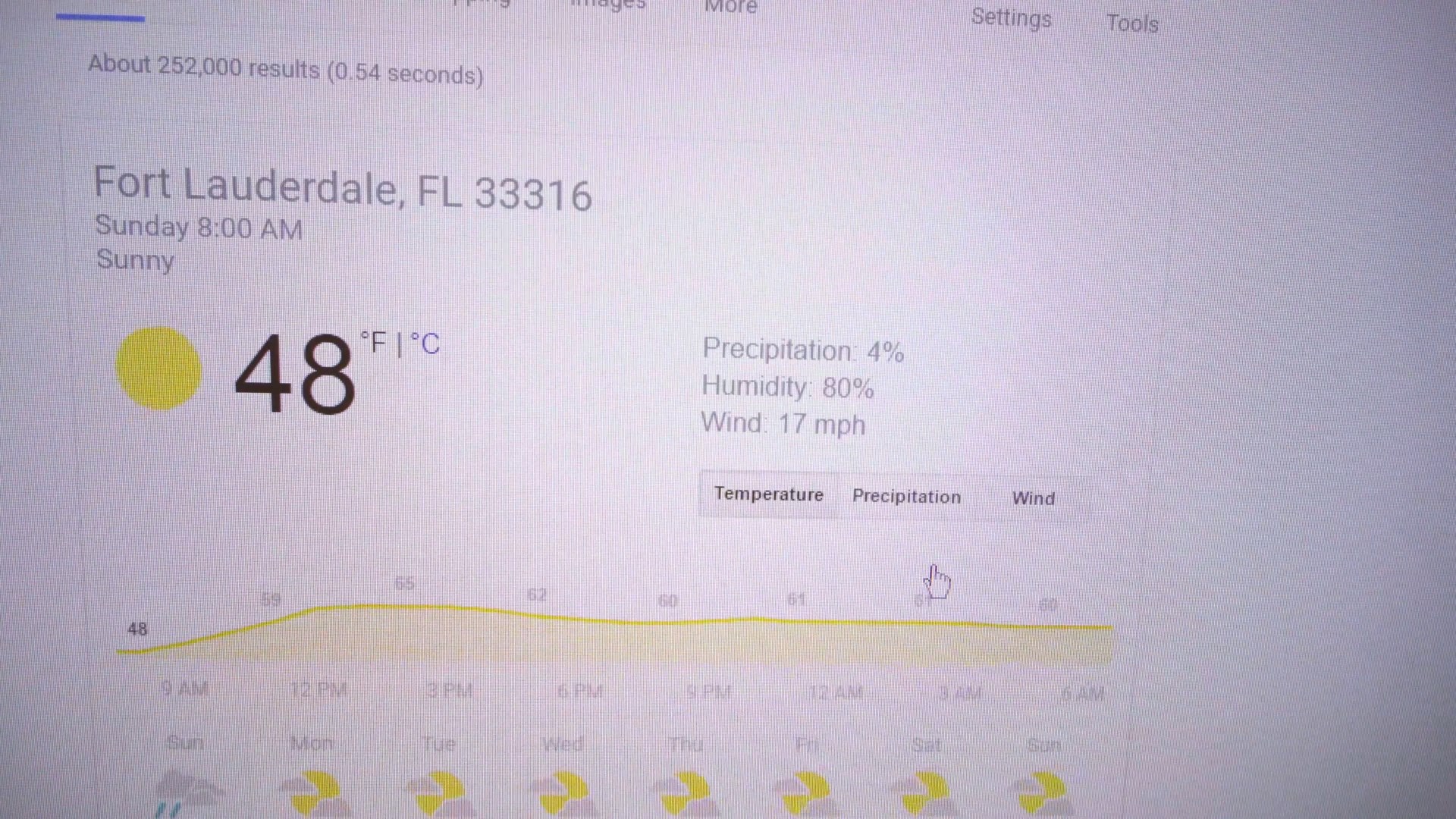}}
\vspace{-3pt}(10) Mixed
\end{minipage}
\begin{minipage}[b]{0.24\linewidth}
  \centering
  \centerline{\includegraphics[width=\textwidth]{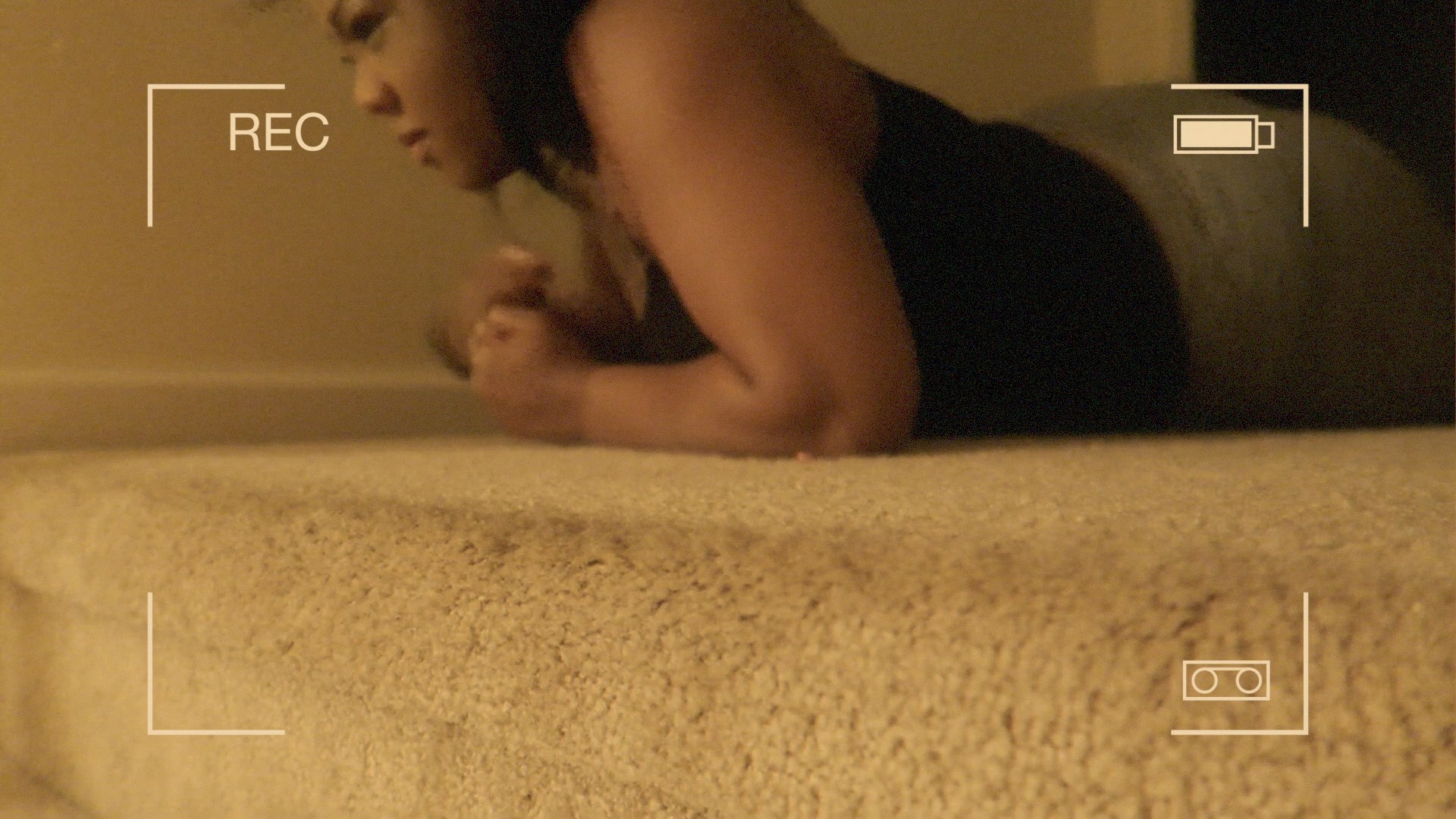}}
\vspace{-3pt}(11) Fitness
\end{minipage}
\begin{minipage}[b]{0.24\linewidth}
  \centering
  \centerline{\includegraphics[width=\textwidth]{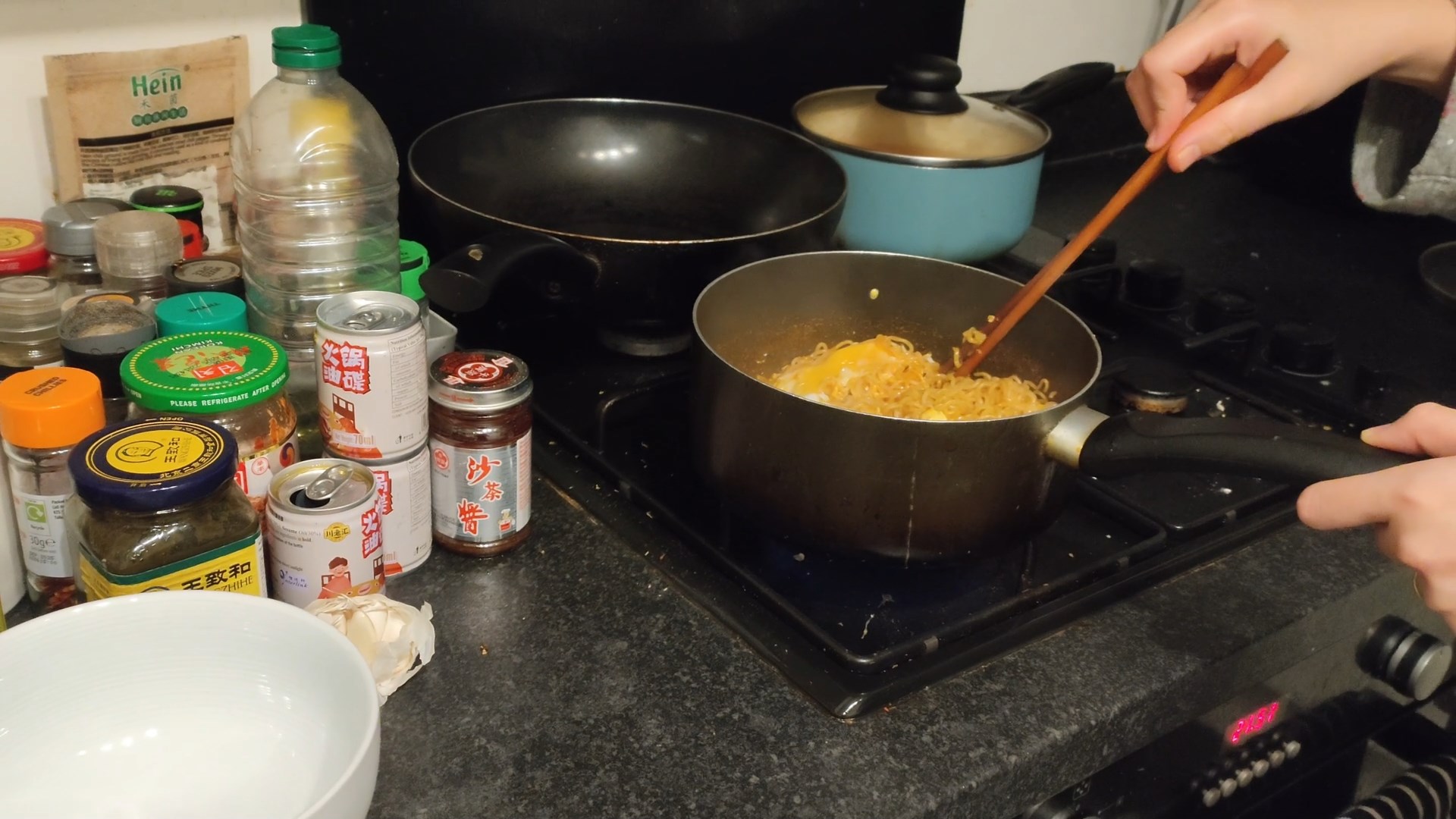}}
\vspace{-3pt}(12) Food
\end{minipage}
\end{minipage}
\hspace{-2pt}\begin{minipage}[b]{0.45\linewidth}
\begin{minipage}[b]{0.32\linewidth}
  \centering
  \centerline{\includegraphics[width=\textwidth]{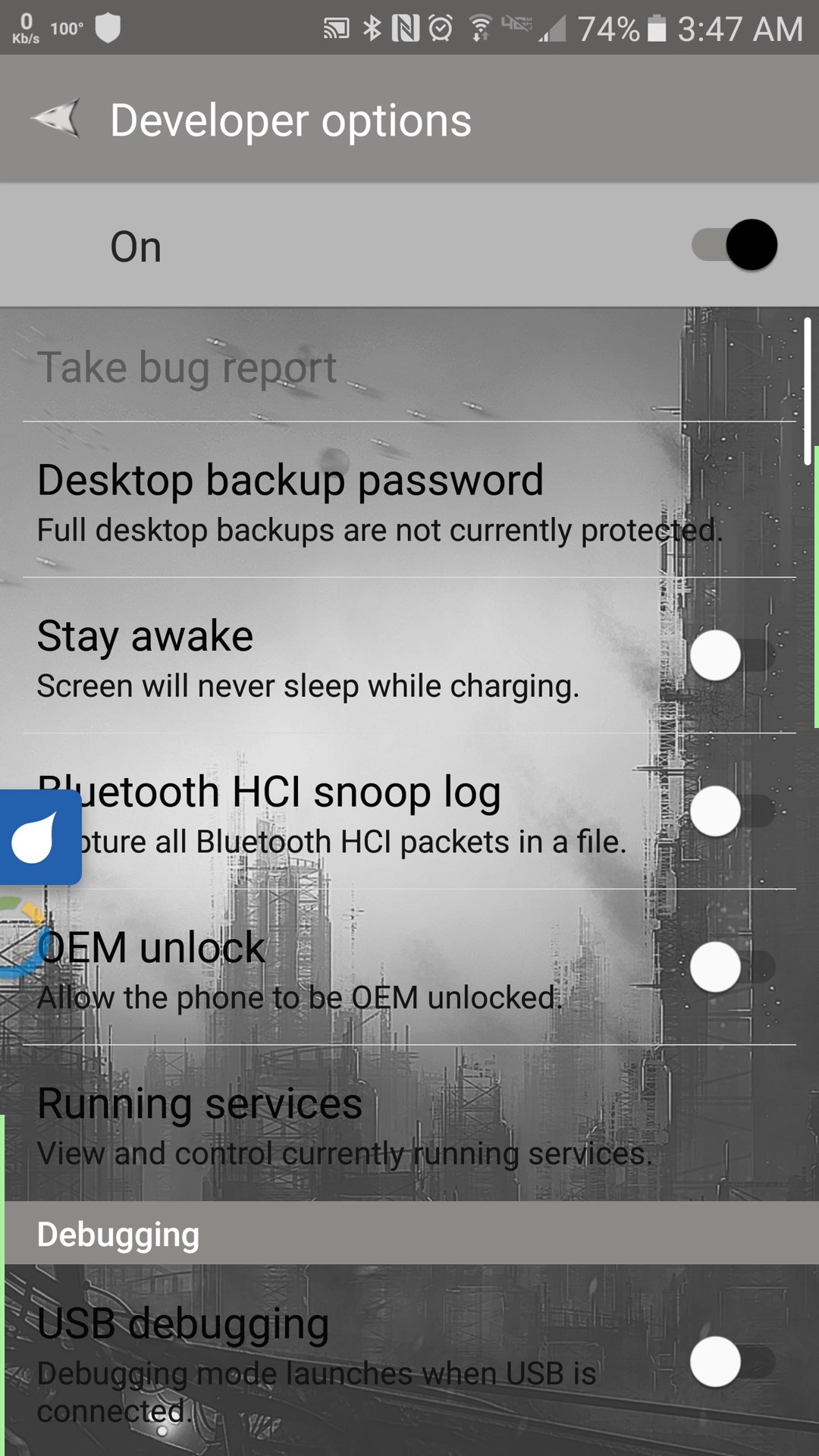}}
\vspace{-3pt}(13) Tech\&DIY
\end{minipage}
\begin{minipage}[b]{0.32\linewidth}
  \centering
  \centerline{\includegraphics[width=\textwidth]{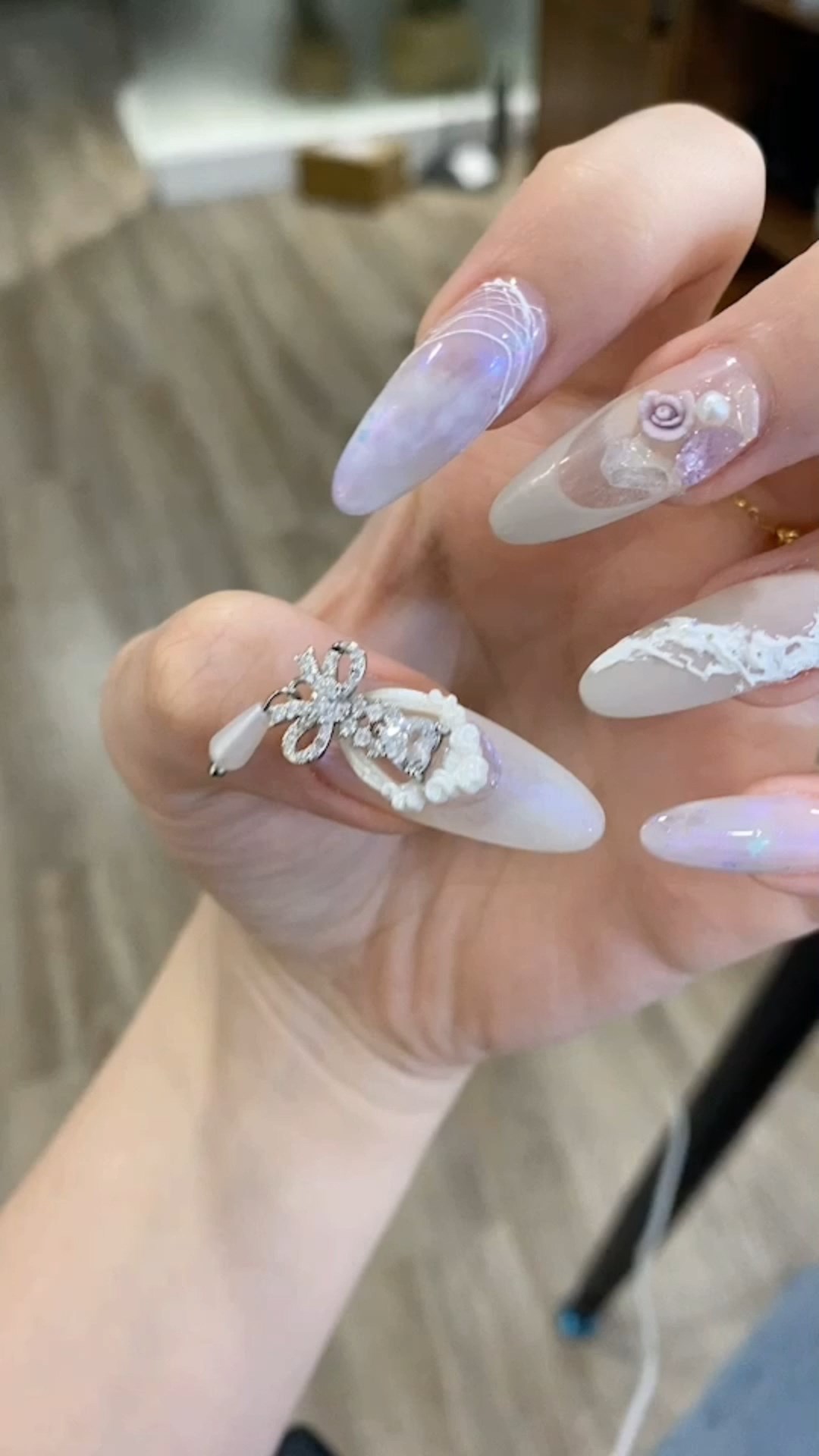}}
\vspace{-3pt}(14) Beauty
\end{minipage}
\begin{minipage}[b]{0.32\linewidth}
  \centering
  \centerline{\includegraphics[width=\textwidth]{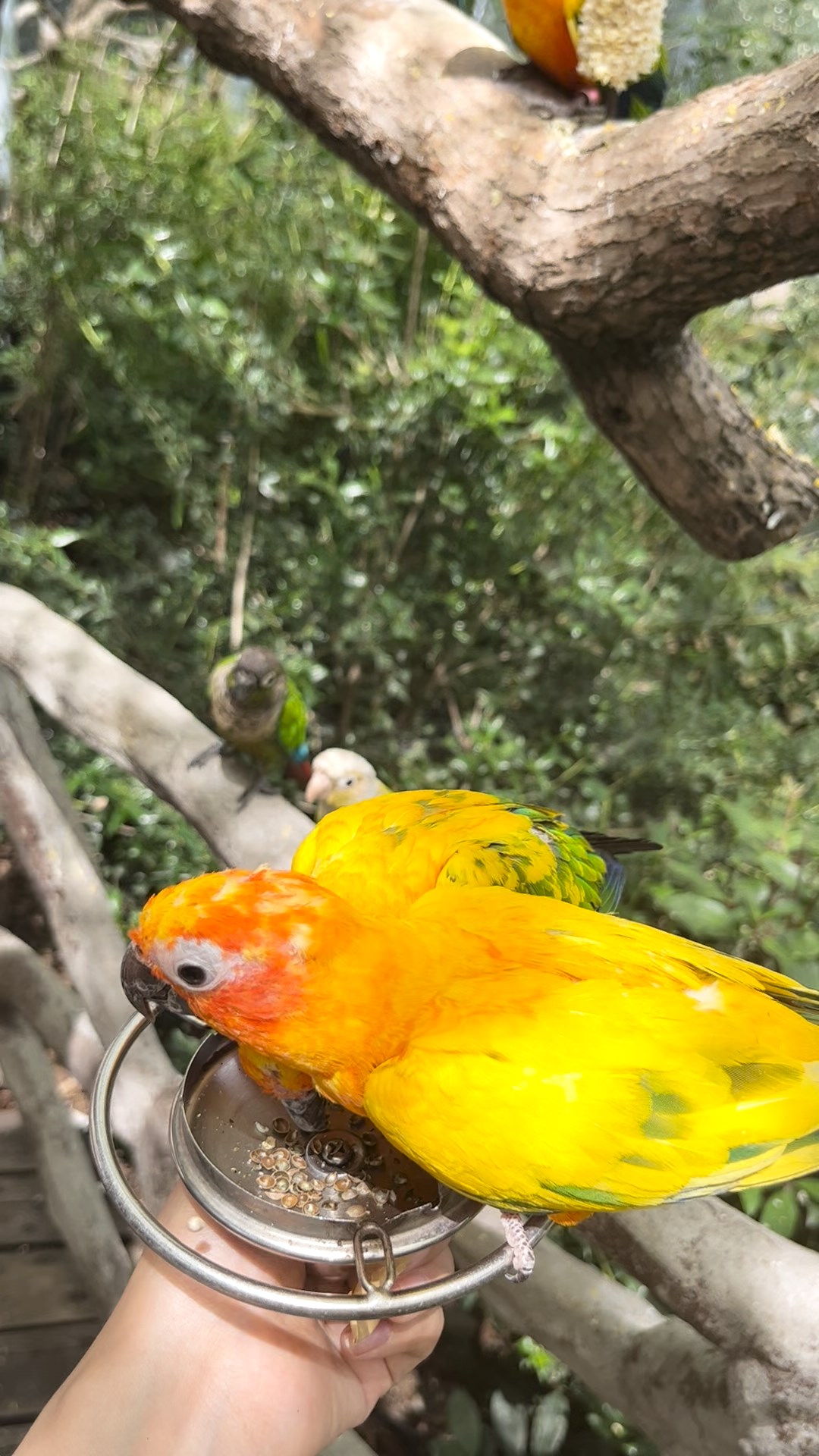}}
\vspace{-3pt}(15) Animal
\end{minipage}
\end{minipage}
\caption{Example frames in selected videos from 15 UGC catogories defined in this database. Every category contains both landscape and portrait videos (the ratio is approximately 2:1). }%To be noted, the sample frames presented are randomly picked from the content. 
\label{fig:sample}
\vspace{-5pt}\end{figure*}

In the context of UGC, most subjective quality assessment approaches focus on the no-reference scenario. Existing publicly available UGC databases such as YouTube UGC~\cite{wang2019youtube}, KoNViD-1k~\cite{konvid1k,hosu2017konstanz}, KonViD-150k~\cite{konvid150k,hahn2021}, and LIVE-VQC~\cite{sinno2018large} only provide distorted (transcoded) sequences in the absence of their corresponding (non-pristine) reference or (pristine) original content. Only a few databases attempt to extend research to the full-reference (transcoded) case. For example, YouTube-UGC~\cite{wang2019youtube} contains a subset that provides VP9 transcoded content together with their low-quality references. The Quality Assessment Grand Challenge of ICME 2021~\cite{wang2021challenge} released a large UGC database containing 7000 transcoded sequences and their 1000 nearly-pristine references. Similarly, TaoLive~\cite{zhang2023md} is another large-scale database that provides high-quality references together with transcoded content. LIVE-WILD~\cite{yu2021predicting} and UGC-VIDEO~\cite{li2020ugc} are also notable works in this area, consisting of test sequences compressed from medium quality level references. The main features of these four databases are summarized in \autoref{tab_dataset}. It is noted that none of these contains references at multiple quality levels, and their transcoding process is often based on a single video codec (except for the UGC-VIDEO database), which does not reflect the diversity of the trascoded UGC videos.

\section{The BVI-UGC Database}\label{sec:pro}

To address the issues mentioned in \autoref{sec:related} regarding the lack of comprehensive video quality databases for UGC transcoding, we have developed a new database, BVI-UGC which, for the first time, specifically considers the UGC transcoding process. We simulated the UGC video delivery pipeline shown in \autoref{fig:transcoding} to generate non-pristine references at different quality levels, and applied another layer of encoding (transcoding) using three commonly used codecs to obtain distorted video sequences.

This section describes the methodology used to select/capture source sequences in the BVI-UGC database, and the workflow to generate non-pristine reference and transcoded content.

\subsection{Source Sequences}

In order to develop a database with diverse and representative source content, we first defined 15 typical UGC categories following the scope of existing UGC databases and the genres on popular UGC platforms (e.g. YouTube, Tiktok,  Flickr etc.) \cite{wang2019youtube}, including (1) Lyric Video; (2) Animation; (3) Game; (4) Streamer; (5) Sports; (6) Travel; (7) Lifestyle; (8) Performance; (9) Lecture; (10) Mixed;  (11) Fitness; (12) Food; (13) Tech \& DIY; (14) Beauty; (15) Animal. In each category, ten videos were collected from the YouTube-UGC database~\cite{wang2019youtube} or captured using mobile phones and drones. The scenes captured in these sequences were designed to diversify low and high level features, such as spatial textures, motions, lighting conditions, backgrounds and foregrounds. This results in a total of 150 ten-second candidate videos. It should be noted that we ensured that all the source sequences are visually lossless so that they can be used as `pristine' original content in the UGC pipeline.

\begin{figure}[t]
\small
\begin{minipage}[b]{0.495\linewidth}
  \centering
  \centerline{\includegraphics[width=1.05\textwidth]{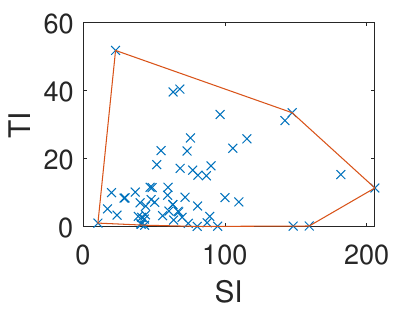}} 
\end{minipage}
\begin{minipage}[b]{0.495\linewidth}
  \centering
  \centerline{\includegraphics[width=1.05\textwidth]{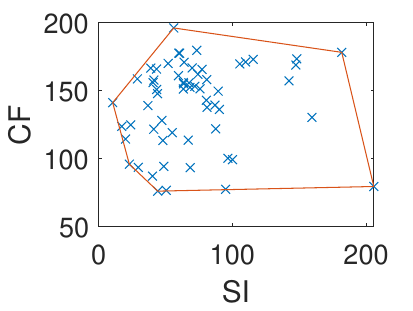}} 
\end{minipage}

\caption{Feature distribution of the source content in the BVI-UGC database. (Left) SI versus TI; (right) SI versus CF.}
\label{fig_coverage}
\vspace{-15pt}\end{figure}

The 150 candidate videos were then truncated into 300 short (5 second) clips, following the optimal duration study conducted in \cite{moss2015optimal,moss2016s}. This allows us to generate more test sequences given the limited time and financial resources. To support further content selection, we follow the procedures in~\cite{zhang2018bvi, ma2021bvi} to determine 60 final source sequences (four for each UGC category). It is also noted that, considering that many UGC videos are associated with a portrait layout, we selected both landscape and portrait videos with a ratio (between landscape and portrait content) of 2:1. Example frames of the selected videos from each of the 15 categories are shown in \autoref{fig:sample}. To showcase the content distribution of the collected content, we calculated these primary video features for each source sequence, including Spatial Information (SI), Temporal Information (TI) and Colorfulness (CF) for each source sequence, following the feature definitions in~\cite{winkler2012analysis}, with the average values of these features (at the sequence level) shown in \autoref{fig_coverage}. It can be observed that the collected content covers a relatively wide range for each video feature compared to other databases in the literature \cite{zhang2018bvi,danier2023bvi,tu2021ugc}.

\begin{figure*}[!t]
\begin{minipage}[b]{1.0\linewidth}
  \centering
  \centerline{\includegraphics[width=\textwidth]{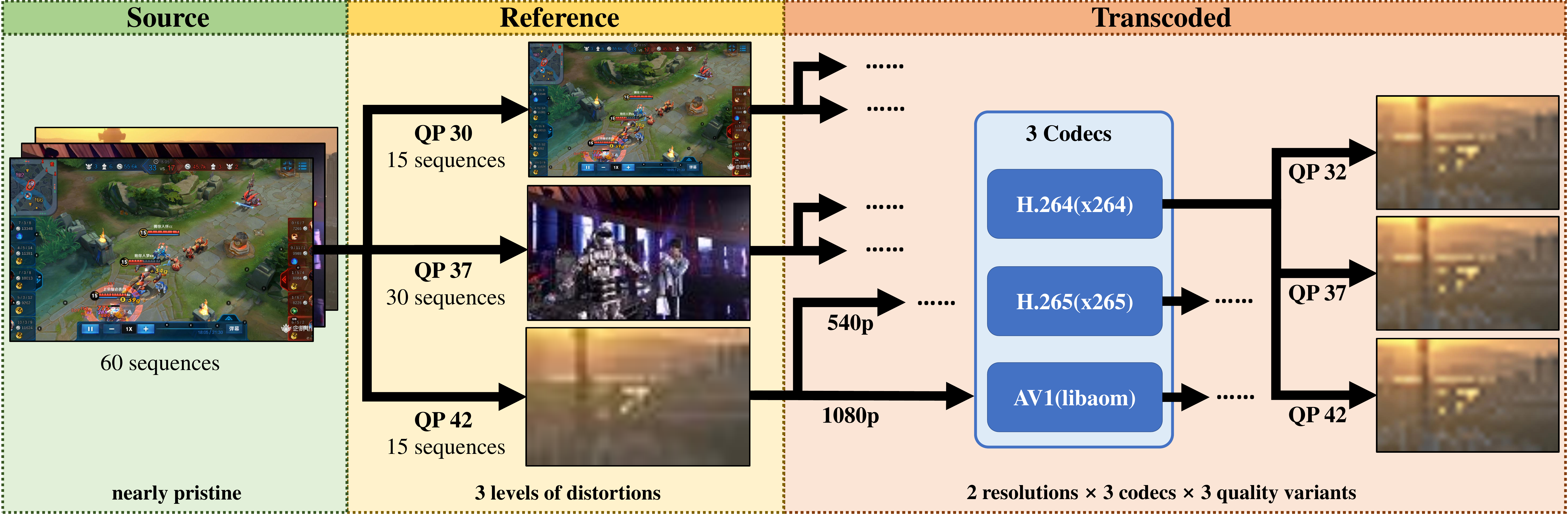}}
\end{minipage}
\caption{Illustration of the content generation process for the BVI-UGC database, which contains 60 non-pristine reference and 1080 transcoded sequences.} % too fat
\label{fig_gen}
\vspace{-5pt}\end{figure*}

\subsection{Non-pristine References}

\begin{figure}[t]
\footnotesize
\begin{minipage}[t]{0.24\linewidth}
\begin{minipage}{\linewidth}
  \centering
  \centerline{\includegraphics[width=\textwidth]{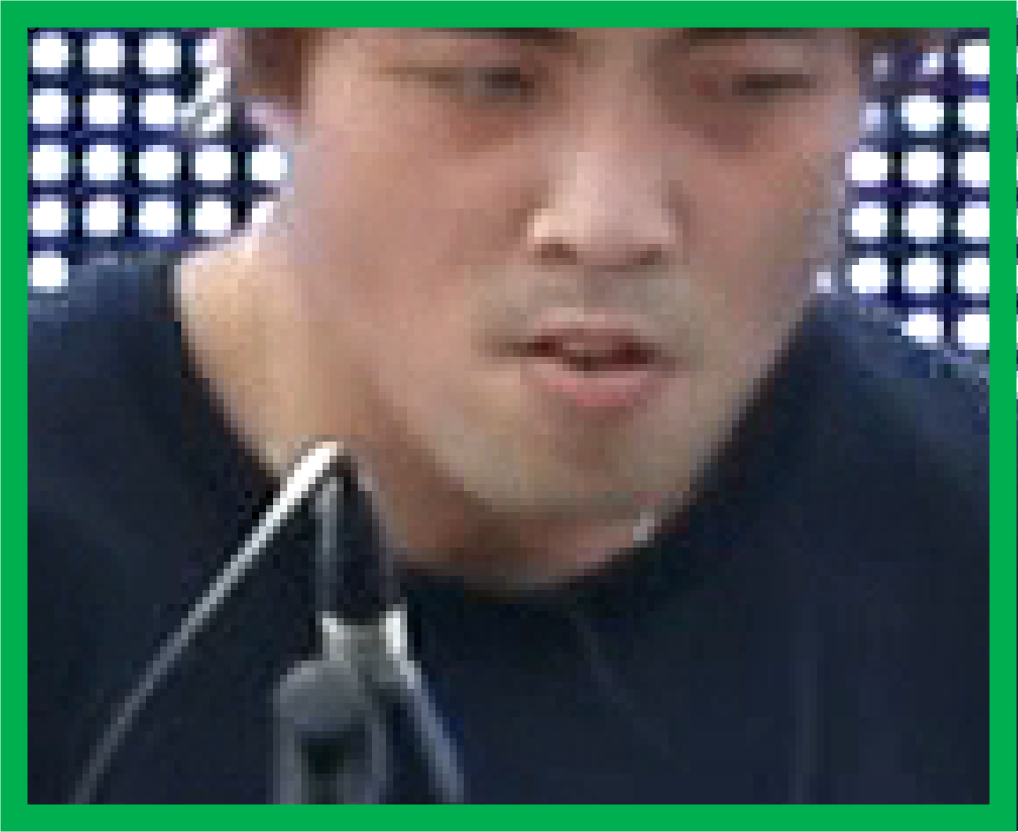}}
\vspace{-3pt}src
\end{minipage}

\begin{minipage}{\linewidth}
  \centering
  \centerline{\includegraphics[width=\textwidth]{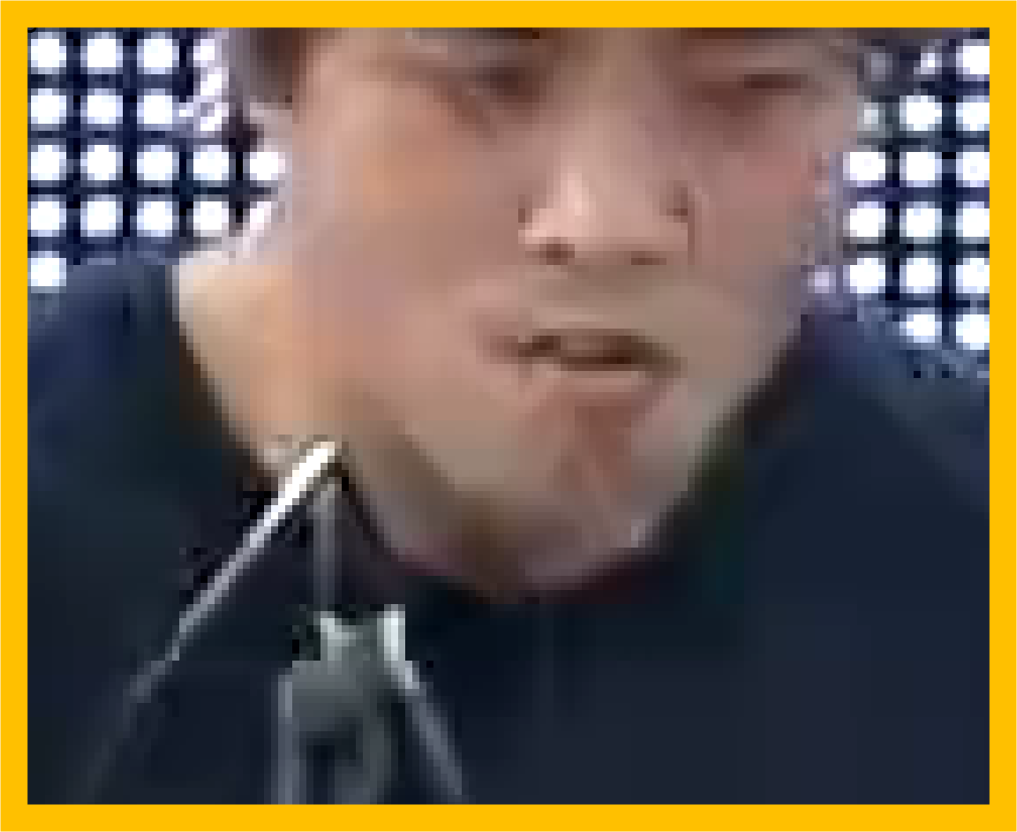}}
\vspace{-3pt}ref (x264:QP42)
\end{minipage}

\begin{minipage}{\linewidth} \centerline{\includegraphics[width=\textwidth]{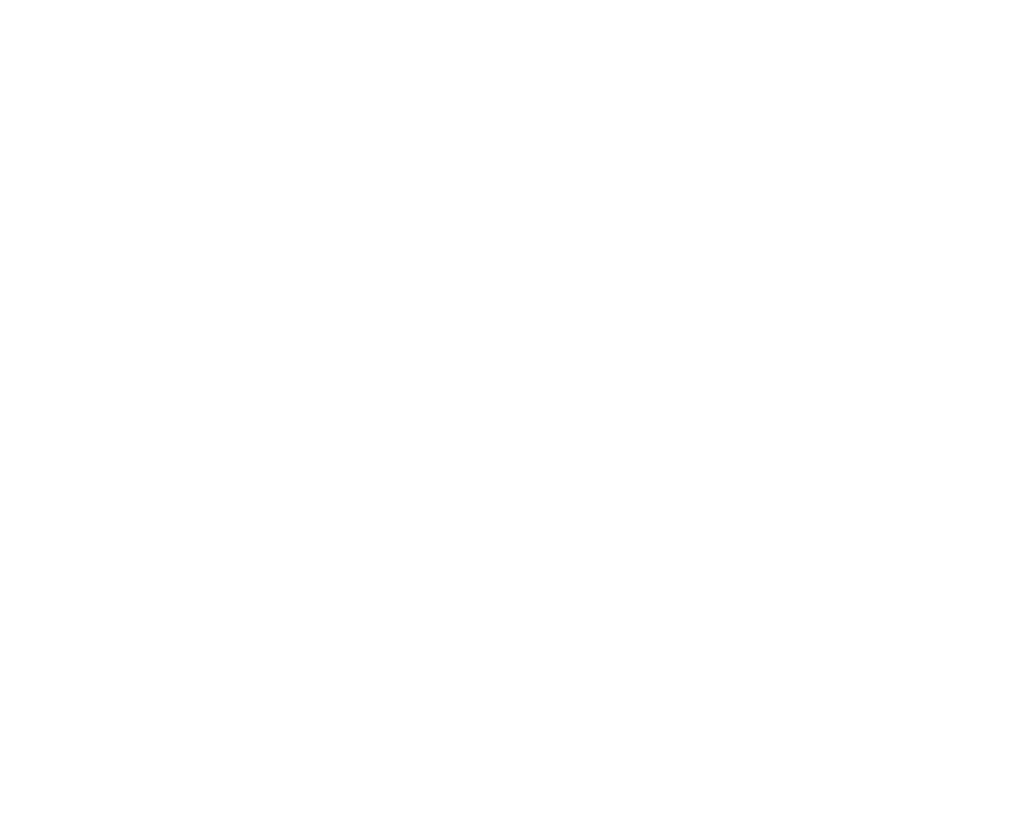}}
\end{minipage}
\end{minipage}
\begin{minipage}[t]{0.745\linewidth}
\begin{minipage}{\linewidth}  \centerline{\includegraphics[width=\textwidth]{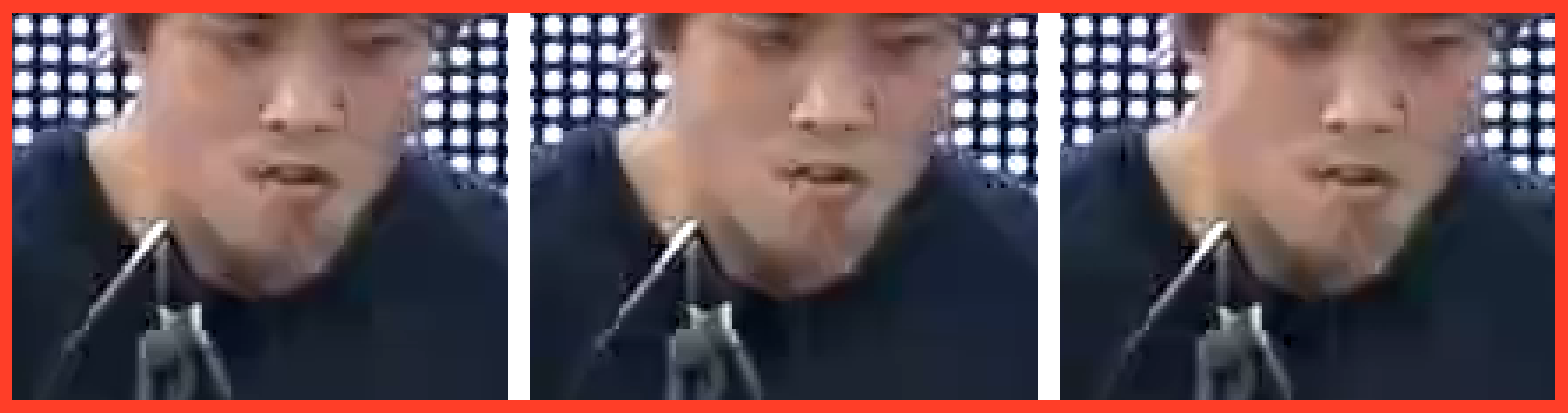}}
\vspace{-3pt}x264: \ \ QP32 \ \ \ \ \ \ \ \ \ \ \ \ QP37 \ \ \ \ \ \ \ \ \ \ \ \ \ \ \  QP42
\end{minipage}

\begin{minipage}{\linewidth}  \centerline{\includegraphics[width=\textwidth]{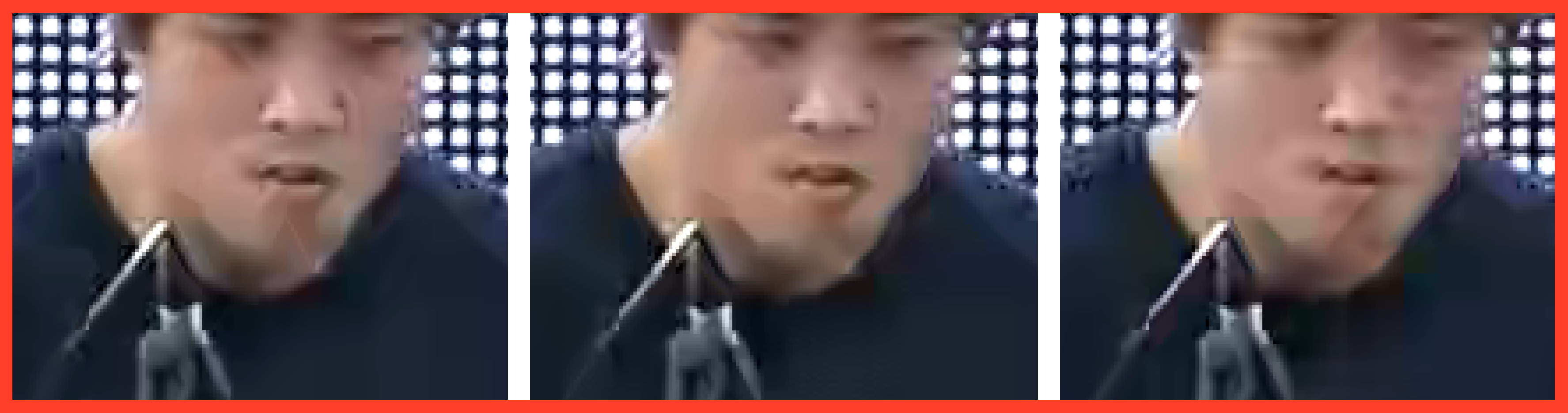}}
\vspace{-3pt}x265: \ \ QP32 \ \ \ \ \ \ \ \ \ \ \ \ QP37 \ \ \ \ \ \ \ \ \ \ \ \ \ \ \  QP42
\end{minipage}

\begin{minipage}{\linewidth}
  \centerline{\includegraphics[width=\textwidth]{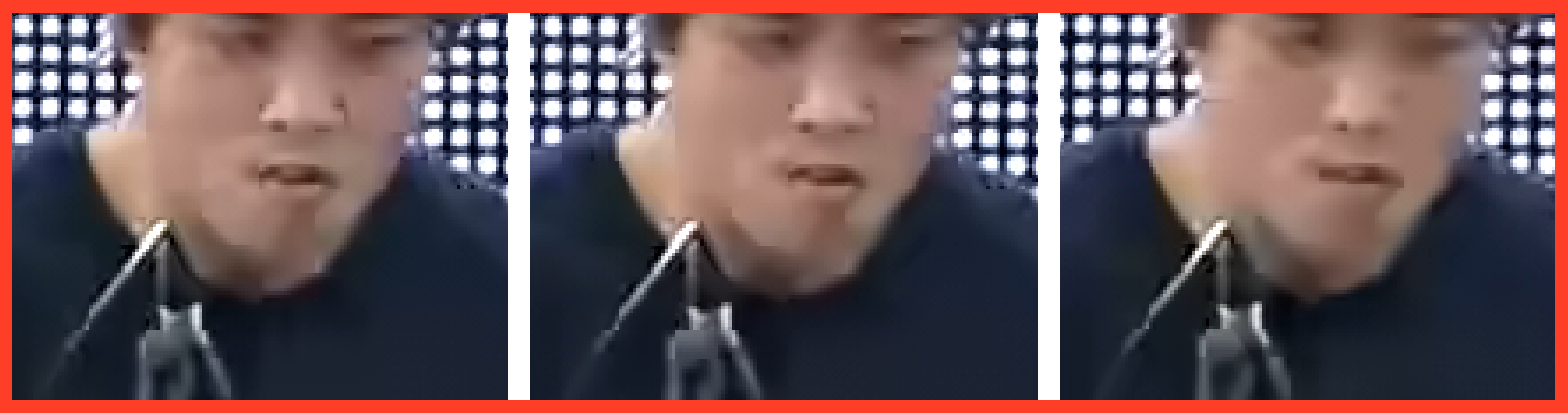}}
\vspace{-3pt}libaom: QP43 \ \ \ \ \ \ \ \ \ \ \ \ QP55 \ \ \ \ \ \ \ \ \ \ \ \ \ \ \  QP63
\end{minipage}
\end{minipage}
\caption{Sample blocks from the high quality source, non-pristine reference and transcoded videos (for the same content).}
\label{fig:compare}
\end{figure}

In a real-world UGC production pipeline, pristine original content is not available on a user's device due to its large storage requirement. Instead, it is typically compressed using a fast video codec to generate compressed sequences that may contain visual artifacts. Their quality may vary, as different coding configurations can be employed here. To simulate this process, we compressed 60 source sequences using a fast implementation of H.264/AVC ~\cite{h264}, x264~\cite{x264repos}, based on its \textit{slow} preset, which is one of the most commonly used video codecs in practical applications. To further diversify the quality of reference sequences, three quantization parameters were used to generate content with high (QP=30), medium (QP=37) and low (QP=42) visual quality levels, as shown in \autoref{fig_gen}. For each source sequence, only one QP value was employed for compression, and we ensured that there is at least one video in every content category encoded with each QP level. This results in 15 non-pristine reference sequences in the QP30 group, 30 in the QP37 group, and 15 in the QP42 group. We generated more non-pristine reference sequences in the medium-quality QP group based on the assumption that this is more common in the real-world UGC production scenarios.

\subsection{Distorted Sequences}\label{sec:datageneration}

Sixty non-pristine references are then compressed again to simulate the transcoding operation. In order to generate more diverse content, we used three video codecs, x264 (ffmpeg v4.4.1 built-in~\cite{ffmpeg}, \textit{slow} preset), x265 (ffmpeg v4.4.1 built-in~\cite{ffmpeg}, \textit{slow} preset), and libaom~\cite{aom2020av2} (v1.0.0, Random Access~\cite{zhao2021aom}), which are commonly employed by various UGC streaming platforms. For each codec, three different QP values are used to produce content at various quality levels. They are \{32, 37 and 42\} for x264 and x265, and \{43, 55 and 63\} for libaom. To further emulate the real-world streaming scenarios, resolution adaptation has also been applied (with a factor of two) in compression. This results in a total number of 1,080 transcoded sequences (60 references $\times$ 2 resolutions $\times$ 3 codecs  $\times$ 3 quality variants, 18 per reference). The content generation workflow is illustrated in \autoref{fig_gen}.

Visual examples have been provided to demonstrate quality differences between high-quality sources, distorted references, and transcoded content, as shown in \autoref{fig:compare}. It can be observed that, when the distorted reference is of low quality (i.e. compressed by x264 using QP42), it is likely that the transcoded sequence will be associated with slightly better perceptual quality due to the artifact filtering and smoothing effect through compression; for example, in \autoref{fig:compare}, the transcoded content by x265 (QP37) looks better than the unpristing reference sequence (x264, QP42). This could challenge FR VQA methods, most of which assume perfect quality with reference content.

%\vspace{-5pt}
\section{Subjective Experiments}
\label{sec:exp}

This section describes the configuration of the crowdsourcing-based subjective experiment, which collected the quality opinion scores of the video sequences in the BVI-UGC database. We then analyze the subjective results and demonstrate its reliability. 

\subsection{Experiment Design}

\begin{algorithm}[t!]
\small
\DontPrintSemicolon
  \SetAlgoLined
  \KwIn {Raw subjective scores $\{o_{ij}\}$, subject index $i \in \{1,..,I_j\}$ and sequence index $ j \in \{1,..,J_i\}$, in which $I_j$ are $J_i$ the number of subjects that have scored sequence $j$ and the number of sequences scored by subject $i$, respectively.}
  \KwOut {Weighted mean opinion score $\mathrm{MOS}_{j}$, standard error of score $\mathrm{SE}_{j}$, subject inconsistency $\mathrm{\sigma}_{i}$ and subject bias $\mathrm{b}_{i}$}
  Initialization\;
  $\mathrm{MOS}_{j,0}  \gets \frac{1}{I_j}\sum^{I_j}_{i=1}o_{ij}$\;
  $\mathrm{b}_{i}  \gets \frac{1}{J_i}\sum^{J_i}_{j=1}(o_{ij} - \mathrm{MOS}_{j,0})$\;
  $t \gets 0$ \;
  \While{$t < 1000$}{
    $r_{ij} = o_{ij} - \mathrm{MOS}_{j,t} - \mathrm{b}_{i}$ \;
    $r_{i} =  \frac{1}{J_i}\sum^{J_i}_{j=1}r_{ij}$ \;
    $\sigma_{i} = \sqrt{\frac{1}{J_i}\sum^{J_i}_{j=1}(r_{ij} - r_{i})^2}$ \;
    $\mathrm{MOS}_{j,t+1} = \sum^{I_j}_{i=1}\sigma_{i}^{-2}(o_{ij}-\mathrm{b}_{i}) / \sum^{I_j}_{i=1}\sigma_{i}^{-2}$ \;
    $\mathrm{b}_{i}  = \frac{1}{J_i}\sum^{J_i}_{j=1}(o_{ij} - \mathrm{MOS}_{j,t+1})$\;
    \If {$\sum^{J}_{j=1}(\mathrm{MOS}_{j,t+1} - \mathrm{MOS}_{j,t})^2 < 10^{-16}$}{
      break \;
    }
    $t = t + 1$ \;
  }
  $r_{j} =  \frac{1}{I_j}\sum^{I_j}_{i=1}r_{ij}$ \;
  $\mathrm{SE}_{j} = \frac{1}{I_j}\sqrt{\sum^{I_j}_{i=1}(r_{ij} - r_{j})^2}$ \;
\caption{The calculation of MOS based on \cite{recommendation2022910}.}
\label{alg_mos}
\end{algorithm}

\begin{figure}[!t]
\hfill
\small
\centering
\begin{minipage}{0.49\linewidth}
  \centering
  \centerline{\includegraphics[width=1.05\linewidth]{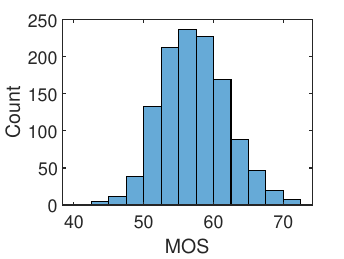}}
\vspace{-3pt}(a)
\end{minipage}
\begin{minipage}{0.49\linewidth}
  \centering
  \centerline{\includegraphics[width=1.05\linewidth]{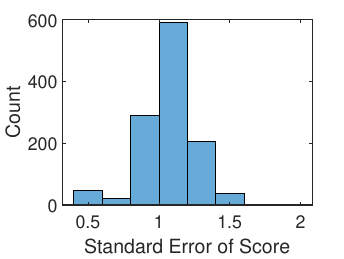}}
\vspace{-3pt}(b) % P.910 document at: https://www.itu.int/rec/T-REC-P.910-202310-I/en
\end{minipage}

\begin{minipage}{0.49\linewidth}
  \centering
  \centerline{\includegraphics[width=1.05\linewidth]{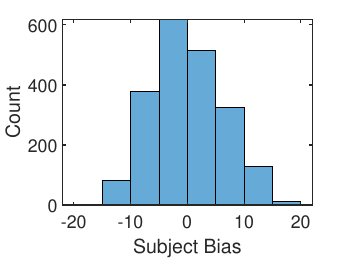}}
\vspace{-3pt}(c)
\end{minipage}
\begin{minipage}{0.49\linewidth}
  \centering
  \centerline{\includegraphics[width=1.05\linewidth]{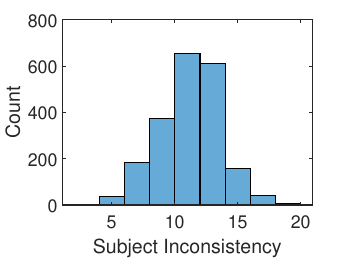}}
\vspace{-3pt}(d)
\end{minipage}
\caption{ Histogram of (a) the MOS; (b) standard error of score over subjects;. (c) subject bias; (d) subject inconsistency. } % change color, maybe pink
\vspace{-10pt}
\label{fig_hist}
\end{figure}

\begin{figure}[htb]
\centering
\begin{minipage}{0.7\linewidth}
  \centering
  \centerline{\includegraphics[width=1.1\textwidth]{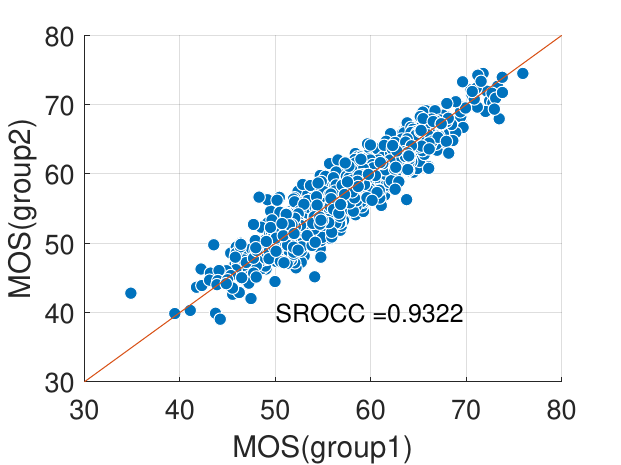}}
\end{minipage}
\caption{Scatter plot of MOS values given by two randomly separated halves of subjective data, from one random split in the 1,000 repetitions. Higher correlated ratio indicates higher inter-subject consistency.}
\label{fig_corr}
\end{figure}

\begin{figure*}[!t]
\hfill
\small
\centering

\begin{minipage}{0.34\linewidth}
  \centering
  \centerline{\includegraphics[width=1.05\linewidth]{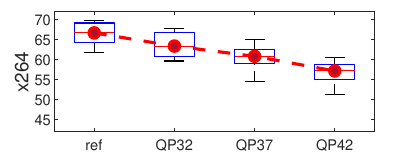}}
%\vspace{-5pt}(a)
\end{minipage}
\hspace{-10pt}\begin{minipage}{0.34\linewidth}
  \centering
  \centerline{\includegraphics[width=1.05\linewidth]{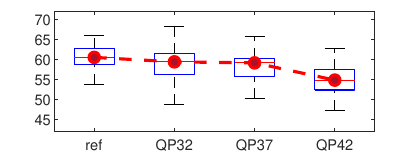}}
%\vspace{-5pt}(b)
\end{minipage}
\hspace{-10pt}\begin{minipage}{0.34\linewidth}
  \centering
  \centerline{\includegraphics[width=1.05\linewidth]{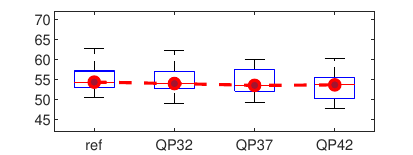}}
%\vspace{-5pt}(c)
\end{minipage}

\begin{minipage}{0.34\linewidth}
  \centering
  \centerline{\includegraphics[width=1.05\linewidth]{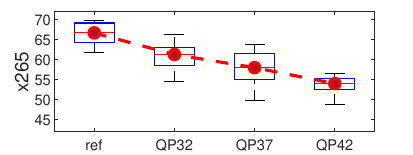}}
%\vspace{-5pt}(d)
\end{minipage}
\hspace{-10pt}\begin{minipage}{0.34\linewidth}
  \centering
  \centerline{\includegraphics[width=1.05\linewidth]{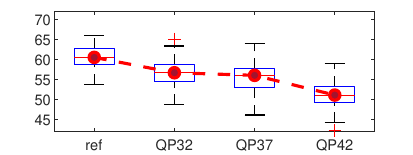}}
%\vspace{-5pt}(e)
\end{minipage}
\hspace{-10pt}\begin{minipage}{0.34\linewidth}
  \centering
  \centerline{\includegraphics[width=1.05\linewidth]{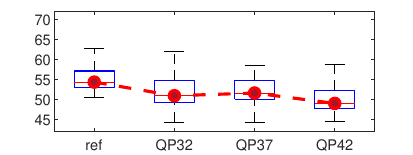}}
%\vspace{-5pt}(f)
\end{minipage}

\begin{minipage}{0.34\linewidth}
  \centering
  \centerline{\includegraphics[width=1.05\linewidth]{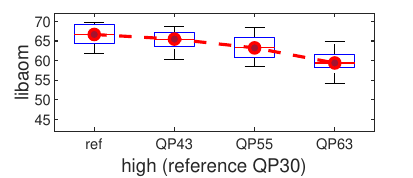}}
%\vspace{-5pt}(g)
\end{minipage}
\hspace{-10pt}\begin{minipage}{0.34\linewidth}
  \centering
  \centerline{\includegraphics[width=1.05\linewidth]{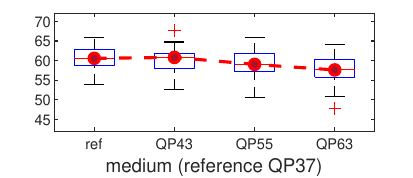}}
%\vspace{-5pt}(h)
\end{minipage}
\hspace{-10pt}\begin{minipage}{0.34\linewidth}
  \centering
  \centerline{\includegraphics[width=1.05\linewidth]{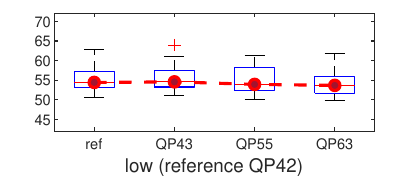}}
%\vspace{-5pt}(i)
\end{minipage}
\caption{Boxplots of MOS of the tested sequences against different quantization parameters during transcoding. Three rows correspond to the three transcoding codecs used. Three columns correspond to the three reference quality groups. } 
%The red lines connect the medium values. Extreme cases are shown with red cross (\textit{Whisker multiplier}=1.5).
\label{fig_boxplot}
\end{figure*}

\begin{figure*}[!t]
\footnotesize
\begin{minipage}[b]{0.245\linewidth}
\centering
\begin{minipage}[b]{0.485\linewidth}
  \centering
  \centerline{\includegraphics[width=\textwidth]{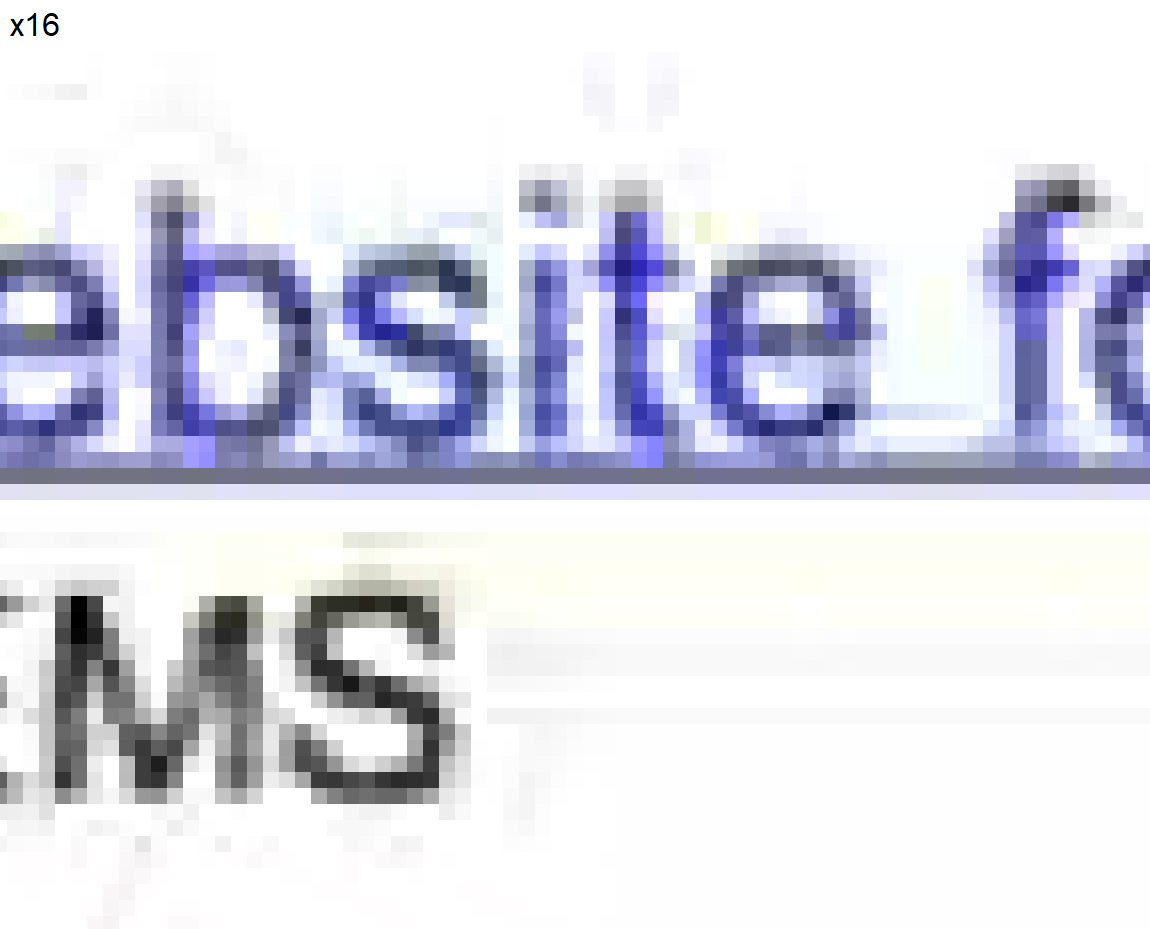}}
\vspace{-3pt}ref\\ (MOS 57.99)
\end{minipage}
\begin{minipage}[b]{0.485\linewidth}
  \centering
  \centerline{\includegraphics[width=\textwidth]{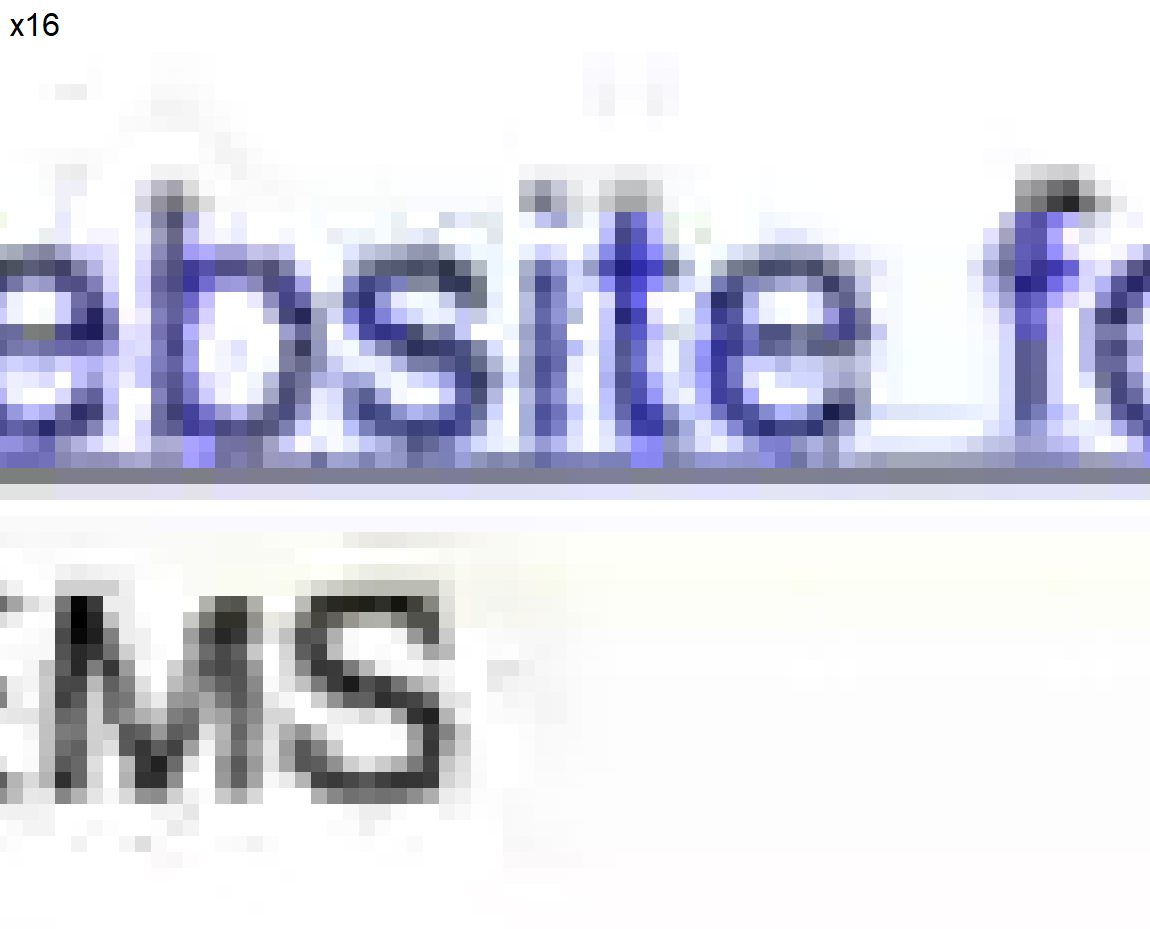}}
\vspace{-3pt}x264-l-QP37\\ (MOS 59.13)
\end{minipage}
\\(1) Lecture7
\end{minipage}
\begin{minipage}[b]{0.245\linewidth}
\centering
\begin{minipage}[b]{0.485\linewidth}
  \centering
  \centerline{\includegraphics[width=\textwidth]{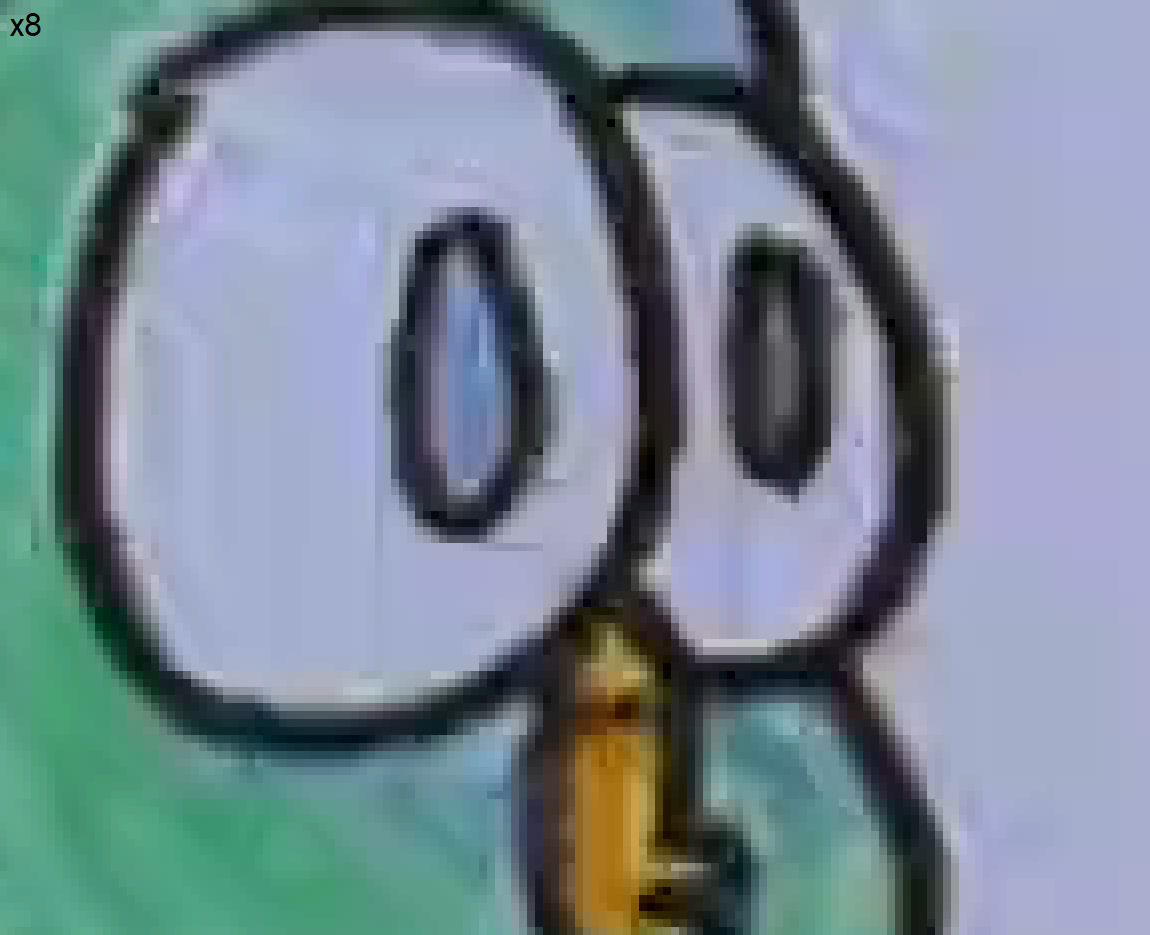}}
\vspace{-3pt}ref\\ (MOS 62.81)
\end{minipage}
\begin{minipage}[b]{0.485\linewidth}
  \centering
  \centerline{\includegraphics[width=\textwidth]{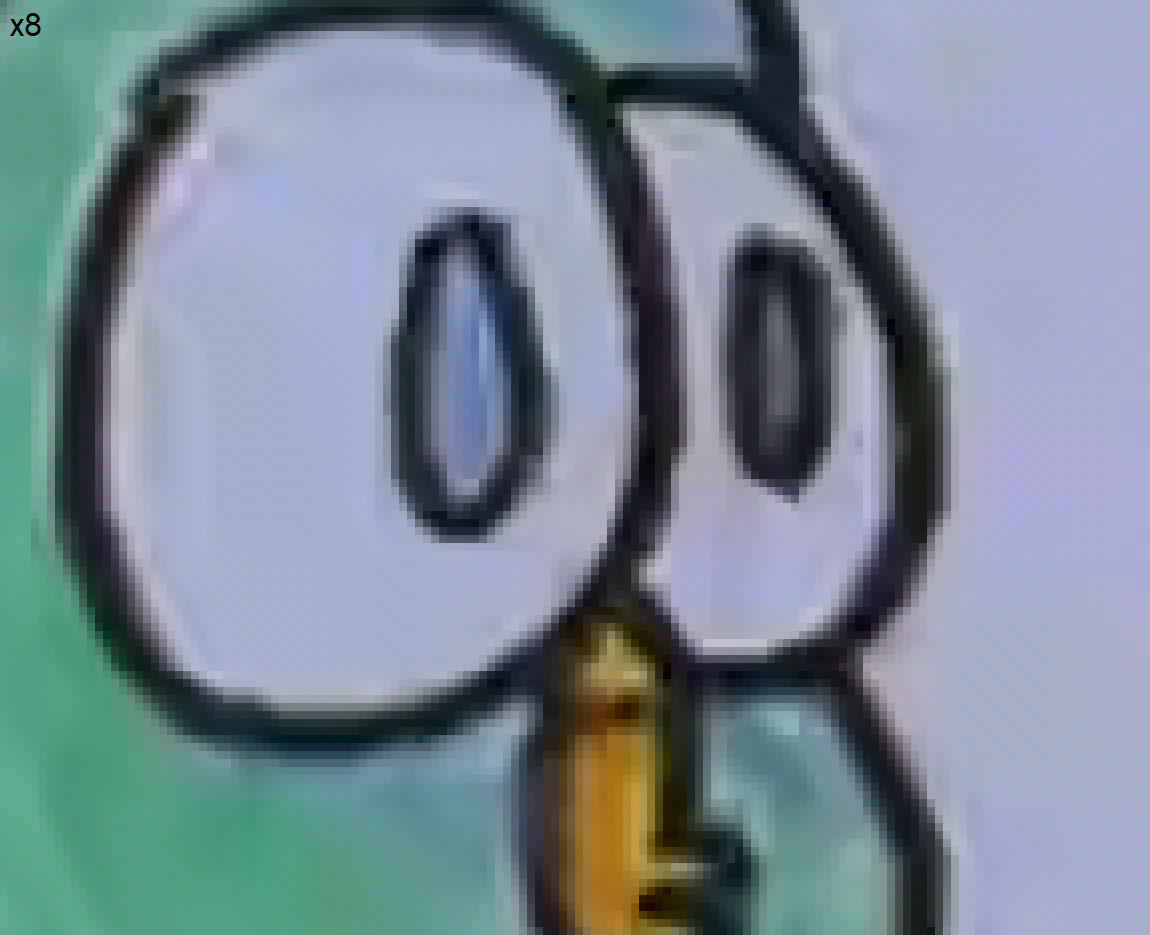}}
\vspace{-3pt}libaom-l-QP43\\ (MOS 63.97)
\end{minipage}
\\(2) Animation6
\end{minipage}
\begin{minipage}[b]{0.245\linewidth}
\centering
\begin{minipage}[b]{0.485\linewidth}
  \centering
  \centerline{\includegraphics[width=\textwidth]{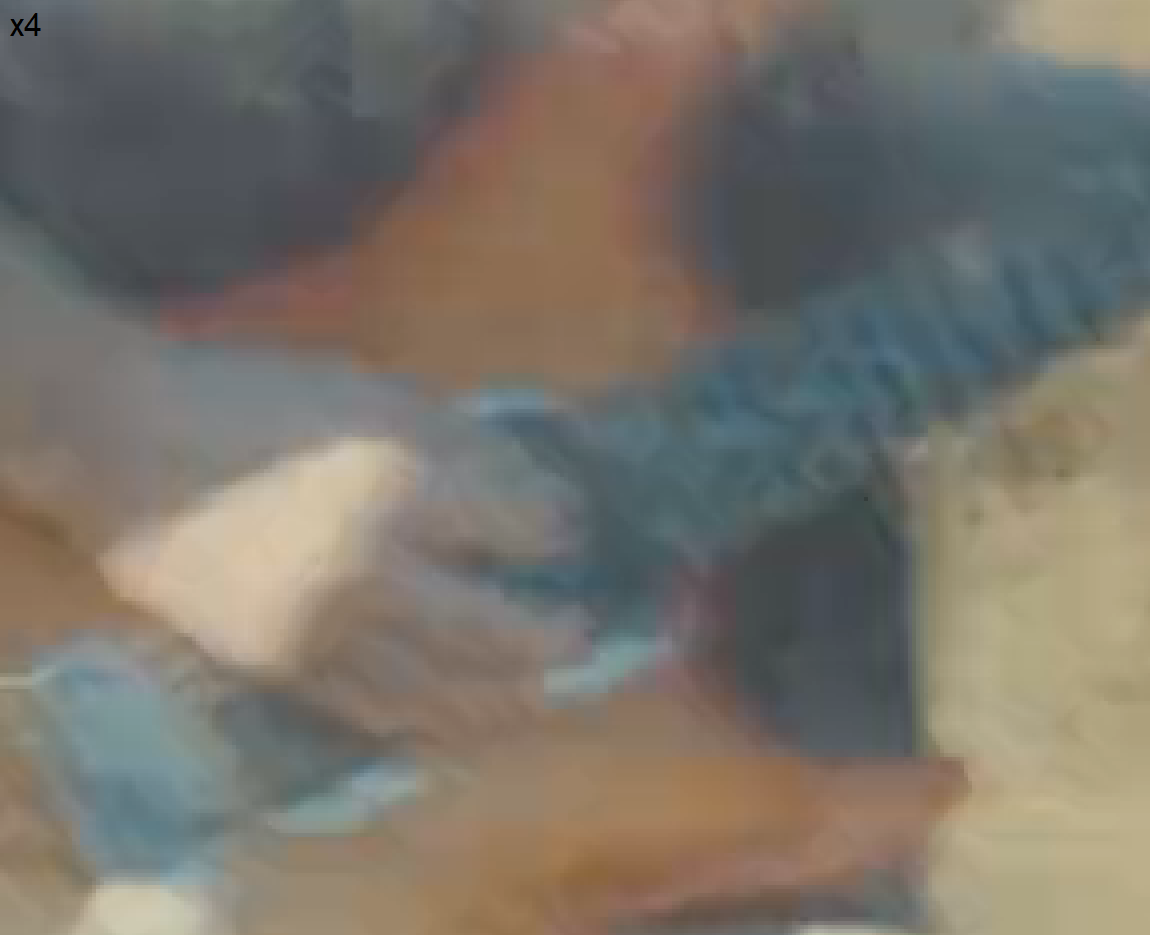}}
\vspace{-3pt}ref\\ (MOS 56.30)
\end{minipage}
\begin{minipage}[b]{0.485\linewidth}
  \centering
  \centerline{\includegraphics[width=\textwidth]{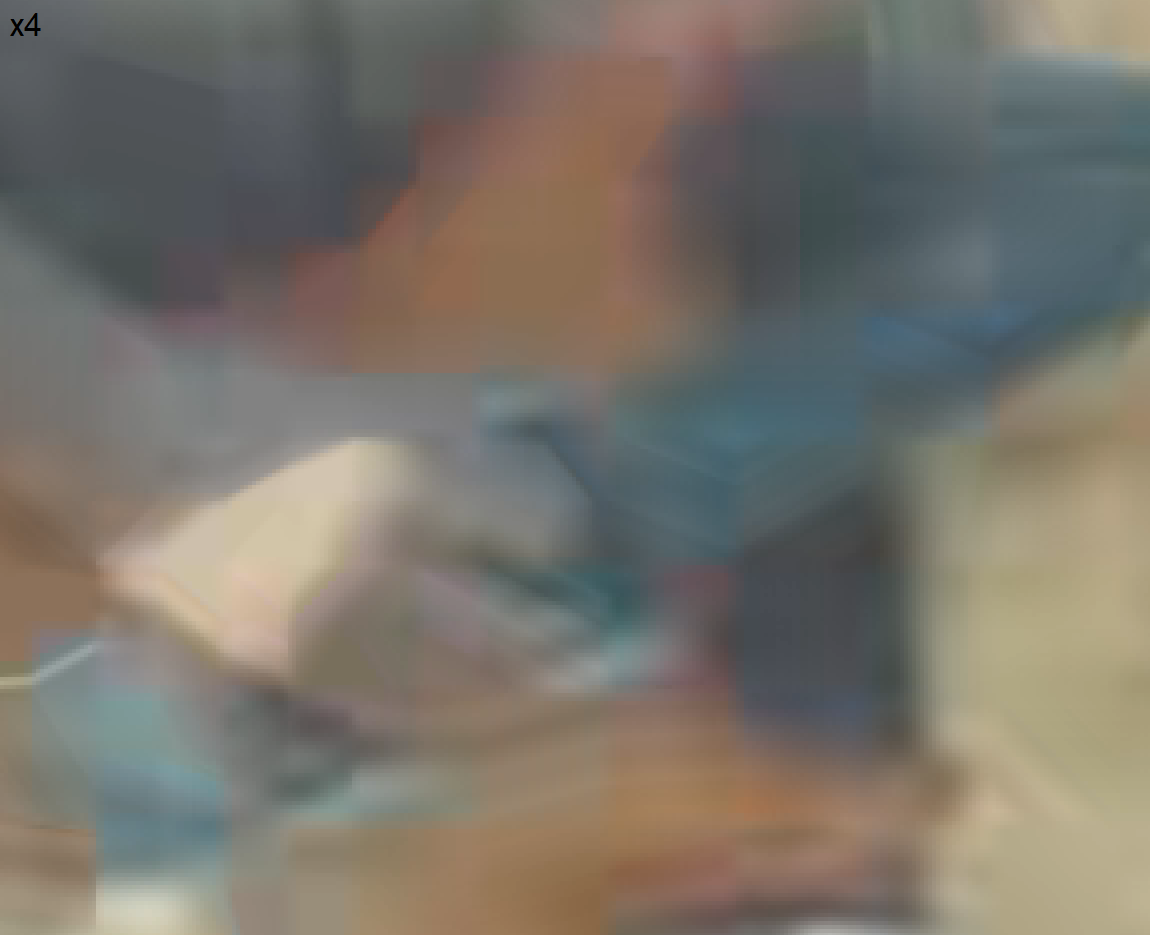}}
\vspace{-3pt}x265-m-QP42\\ (MOS 42.35)
\end{minipage}
\\(3) Lyric15
\end{minipage}
\begin{minipage}[b]{0.245\linewidth}
\centering
\begin{minipage}[b]{0.485\linewidth}
  \centering
  \centerline{\includegraphics[width=\textwidth]{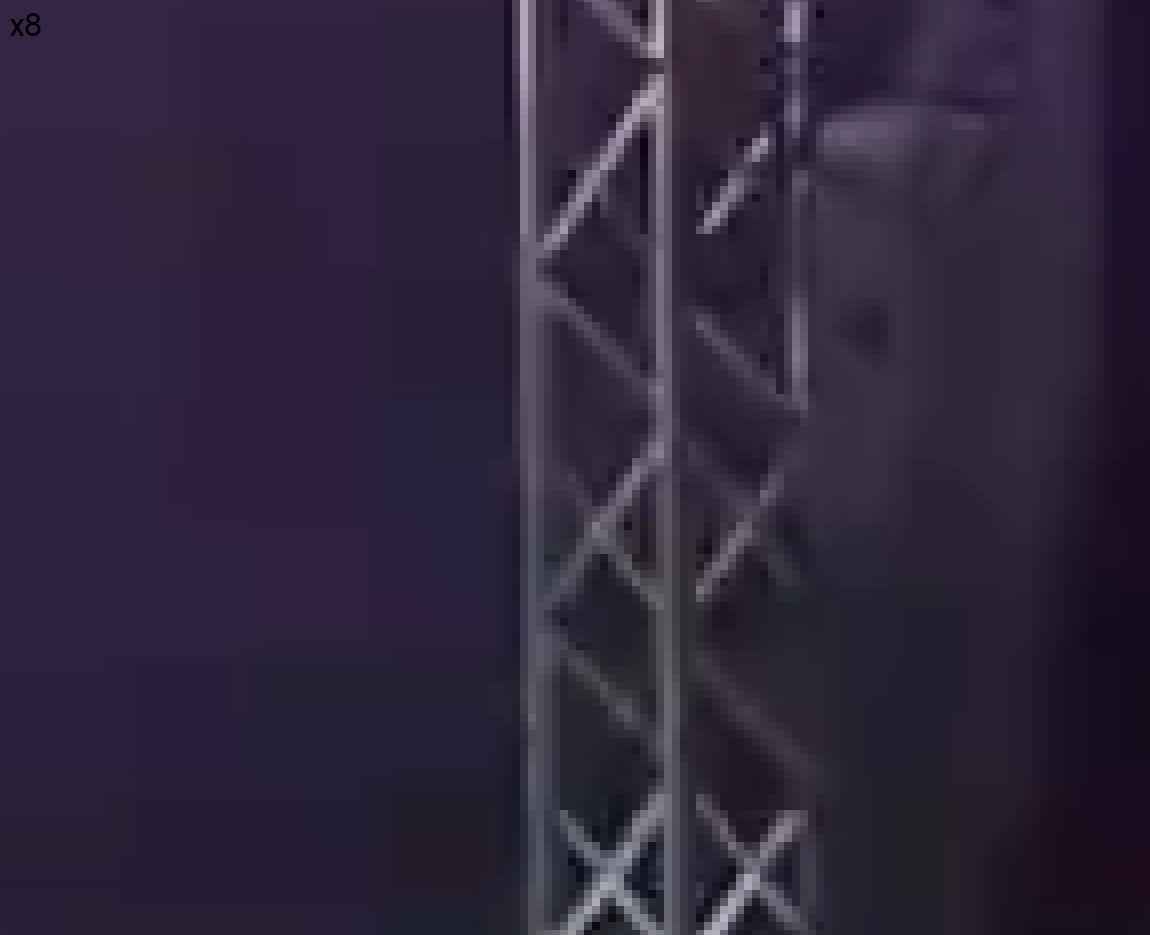}}
\vspace{-3pt}ref\\ (MOS 55.82)
\end{minipage}
\begin{minipage}[b]{0.485\linewidth}
  \centering
  \centerline{\includegraphics[width=\textwidth]{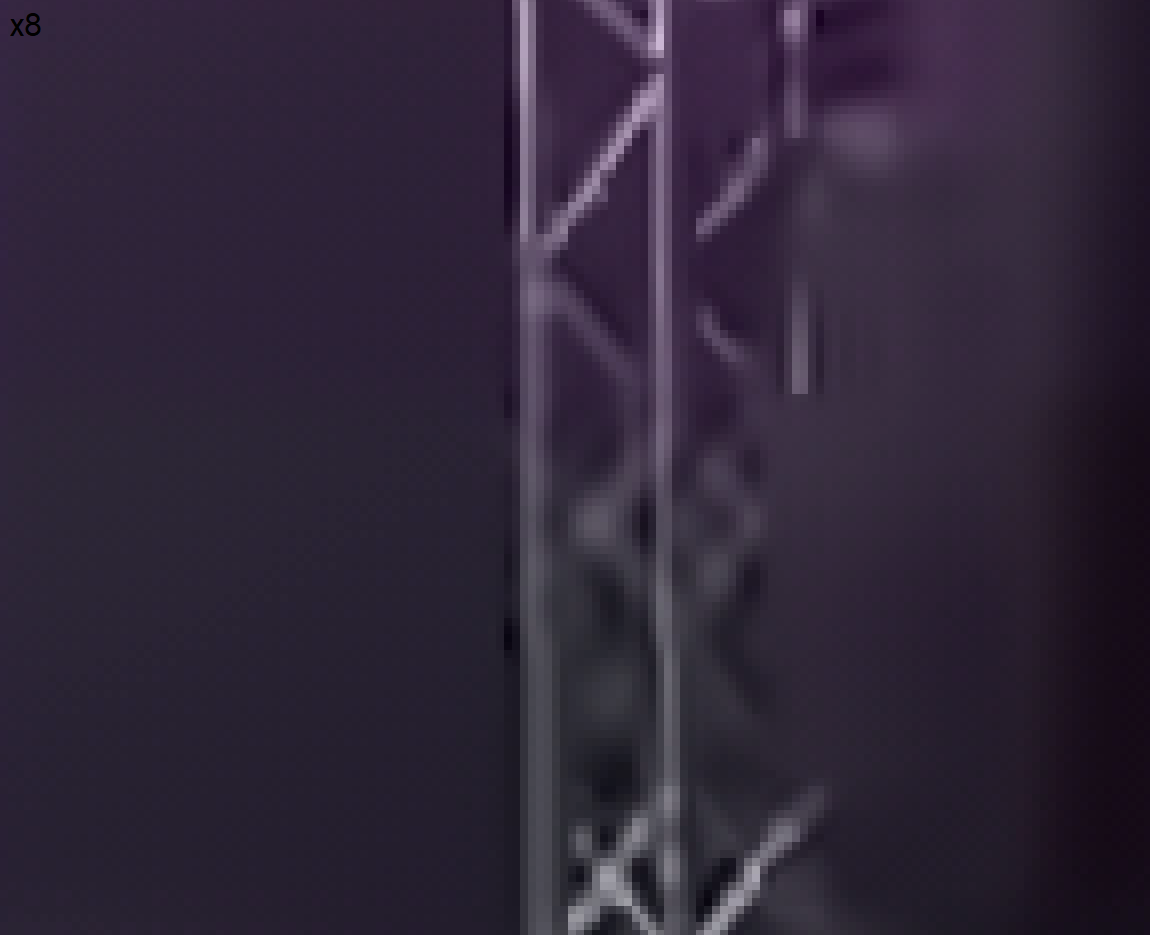}}
\vspace{-3pt}libaom-m-QP63\\ (MOS 47.81)
\end{minipage}
\\(4) Perform3
\end{minipage}
\caption{Visual comparison examples between reference and transcoded content. The labels below figures show the corresponding \textit{codec-group-QP}. For example, x265-m-QP32 means this sample comes from a sequence in the medium reference group, transcoded with x265 codec using QP32.}
\label{fig_outlier}
\end{figure*}

Due to the large number of video sequences in this database and the nature of UGC consumption (typically within inconsistent viewing conditions and on different devices), we designed a crowdsourcing subjective test based on the Amazon Mechanical Turk~\cite{turk2023amazon} platform. In this experiment, we employed the Absolute Category Rating with Hidden Reference (ACR-HR) methodology~\cite{bt2002methodology}, which has been commonly used for many subjective UGC studies~\cite{wang2019youtube,wang2021challenge,yu2021predicting,li2020ugc}, to collect subjective scores for 60 non-pristine reference and 1,080 transcoded sequences. High-quality source content was not shown in this experiment, but is provided alongside other sequences in the database.

In each test session, prior to the formal test, subjects were first asked to calibrate their screen resolution based on the method developed in~\cite{reso_web} and then perform a vision acuity test using Ishihara and Snellen charts. If participants fail in this test, the session is stopped without showing any video sequences. For those who have passed the test, they are shown video sequences randomly picked from 60 non-pristine reference and 1,080 transcoded videos, each of which is played only once. After viewing each video, participants are asked to provide a subjective score of video quality using a continuous slider with five evenly spaced intervals labeled \textit{Bad}, \textit{Poor}, \textit{Fair}, \textit{Good} and \textit{Excellent}, each of which covers a quality range of 20 respectively. To allow subjects to familiarize with the experiment, 3 training trials are presented before the formal test. During the test, participants are free to leave anytime, but the results collected from participants who viewed less than 20 videos were discarded to make sure sufficient videos are watched by each subject.

In this experiment, over 3,500 human participants were paid to provide subjective quality scores, each of whom viewed 53 sequences on average. This ensures more than 160 raw subjective scores collected for each test (non-pristine reference or transcoded) sequence.

\subsection{Data Processing, Validation and Analysis}

Due to the nature of crowdsourcing experiments, subjective data acquired in this experiment may be associated with larger variances. Based on the recommendation in~\cite{recommendation2022910}, we further improved the data reliability by soft screening the collected raw opinion scores when calculating the Mean Opinion Score (MOS), as shown in \hyperref[alg_mos]{Algorithm 1}. The histograms of the resulting MOS values, the corresponding standard errors ($\mathrm{SE}$), subject inconsistency ($\sigma$) and the subject bias ($\mathrm{b}$) are plotted in \autoref{fig_hist}.

To further validate the reliability of the subjective data collected, based on previous works~\cite{madhusudana2021subjective, hossfeld2013best, danier2023bvi}, the raw subjective scores collected for each sequence were randomly divided into two equal groups. For each group, one MOS is calculated for each sequence based on \hyperref[alg_mos]{Algorithm 1}. We then calculated the Spearkman Ranked Order Correlation Coefficient (SROCC) between two groups of MOS for the whole database. This partitioning process has been repeated for 1,000 times to obtain the average SROCC, which is 0.9322. We also provide the scatter plot for one partition in \autoref{fig_corr}. 

Based on the MOS obtained, we can study the influence of reference distortions and different codecs on subjective perceptual experience. We plot the MOS of the transcoded sequences against different quantization parameters for different reference quality levels and transcoding codecs in \autoref{fig_boxplot}. Here, we only focus on the HD content encoding without resolution adaptation. It can be observed that when the reference content is associated with relatively high quality (i.e. reference QP30), applying different QP values during transcoding can generate results with more distinct visual quality levels. In contrast, when the quality of the reference content is low (reference QP42), the visual difference between transcoded content (based on various QP values) is much smaller. We have also provided visual comparison examples between reference and transcoded content in \autoref{fig_outlier}, where in cases (1) and (2), the artifact around the edges were smoothed out during transcoding, which actually improves the visual experience slightly.

\section{Evaluation of Objective Quality Metrics}
\label{sec:metrics}

\begin{table*}[t]
\caption{The correlation results with DMOS of the tested full-/no-reference metrics on BVI-UGC across different reference and codec groups. For no-reference metrics, the difference between quality indices of transcoded video and its non-pristine reference is taken to correlate with DMOS. In each group, the best and the second best metrics are \textbf{boldfaced} and \underline{underlined}.}
\centering
\resizebox{\linewidth}{!}{
\begin{tabular}{r||l|l||l|l||l|l||l|l||l|l||l|l||l|l}
\toprule
\multirow{2}{*}{\textbf{Databases}} &  \multicolumn{2}{c||}{\textbf{BVI-UGC}} &  \multicolumn{2}{c||}{\textbf{BVI-UGC \textcolor{red!85!black}{low}}} &  \multicolumn{2}{c||}{\textbf{BVI-UGC \textcolor{myyellow}{medium}}} &  \multicolumn{2}{c||}{\textbf{BVI-UGC \textcolor{green!70!black}{high}}} &  \multicolumn{2}{c||}{\textbf{BVI-UGC H264}} &  \multicolumn{2}{c||}{\textbf{BVI-UGC H265}} &  \multicolumn{2}{c}{\textbf{BVI-UGC libaom}} \\
\cmidrule{2-15} 
& \textbf{SROCC} & \textbf{KRCC} & \textbf{SROCC} & \textbf{KRCC} & \textbf{SROCC} & \textbf{KRCC} & \textbf{SROCC} & \textbf{KRCC} & \textbf{SROCC} & \textbf{KRCC} & \textbf{SROCC} & \textbf{KRCC} & \textbf{SROCC} & \textbf{KRCC} \\
\midrule \multicolumn{1}{l}{\textbf{Full-reference }} \\
\midrule
PSNR & 0.5003 & 0.3438 & 0.1913 & 0.1279 & 0.5478 & 0.3795 & 0.5459 & 0.3827 & 0.5335 & 0.3694 & 0.4957 & 0.3409 & 0.5731 & 0.4016  \\
SSIM & 0.4663 & 0.3212 & 0.1978 & 0.1328 & 0.5453 & 0.3766 & 0.7004 & 0.5086 & 0.5235 & 0.3644 & 0.5222 & 0.3629 & 0.4733 & 0.3283  \\
ST-GREED~\cite{madhusudana2021st} & 0.1797 & 0.1307 & 0.1818 & 0.1254 & 0.2237 & 0.1626 & 0.1901 & 0.1325 & 0.0857 & 0.0629 & 0.1197 & 0.0916 & 0.3385 & 0.2388  \\
LPIPS~\cite{zhang2018unreasonable} & 0.1398 & 0.0985 & 0.1731 & 0.1231 & 0.1828 & 0.1299 & 0.1444 & 0.0822 & 0.0564 & 0.0374 & 0.0833 & 0.0580 & 0.3207 & 0.2244  \\
VMAF 0.6.1~\cite{li2016toward} & \underline{0.5610} & \underline{0.3903} & \underline{0.4340} & \underline{0.2980} & \textbf{0.6538} & \textbf{0.4604} & \underline{0.7588} & \underline{0.5571} & \textbf{0.6392} & \textbf{0.4521} & \underline{0.6258} & \underline{0.4381} & \underline{0.5948} & \underline{0.4184}  \\
C3DVQA~\cite{xu2020c3dvqa} & 0.4360 & 0.2975 & 0.2678 & 0.1819 & 0.4523 & 0.3035 & 0.5816 & 0.4053 & 0.4810 & 0.3301 & 0.4716 & 0.3194 & 0.4573 & 0.3154  \\
FR-CUGCVQA~\cite{li2021full} & 0.4126 & 0.2836 & 0.1837 & 0.1261 & 0.4914 & 0.3334 & 0.6495 & 0.4597 & 0.5456 & 0.3825 & 0.4986 & 0.3475 & 0.4434 & 0.3120  \\
FR-CONTRIQUE~\cite{madhusudana2022image} & 0.2150 & 0.1561 & 0.1179 & 0.0889 & 0.1544 & 0.1186 & 0.2686 & 0.1916 & 0.6921 & 0.4924 & 0.4820 & 0.3290 & 0.3725 & 0.2552  \\
RankDVQA~\cite{feng2024rankdvqa} & 0.5527 & 0.3842 & 0.1405 & 0.0971 & 0.5118 & 0.3514 & 0.7502 & 0.5567 & 0.5920 & 0.4244 & 0.6055 & 0.4283 & 0.5714 & 0.4080  \\
RankDVQA-UGC~\cite{qi2023full} & \textbf{0.5727} & \textbf{0.4042} & \textbf{0.4420} & \textbf{0.3037} & \underline{0.6478} & \underline{0.4574} & \textbf{0.7617} & \textbf{0.5663} & \underline{0.6347} & \underline{0.4513} & \textbf{0.6351} & \textbf{0.4403} & \textbf{0.5971} & \textbf{0.4193}  \\
\midrule \multicolumn{5}{l}{\textbf{No-reference }(measuring quality degradation, to correlate with DMOS)} \\ 
\midrule
NIQE~\cite{mittal2012making} & 0.1894 & 0.1247 & 0.2381 & 0.1591 & 0.2589 & 0.1714 & 0.1979 & 0.1301 & 0.2099 & 0.1419 & 0.0236 & 0.0165 & 0.2439 & 0.1642  \\
BRISQUE~\cite{mittal2012no} & 0.1316 & 0.0854 & \underline{0.2516} & \underline{0.1628} & 0.0548 & 0.0364 & 0.0824 & 0.0630 & 0.1144 & 0.0742 & 0.1334 & 0.0872 & 0.2210 & 0.1430  \\
VBLIINDS~\cite{saad2014blind} & 0.0651 & 0.0446 & 0.0499 & 0.0367 & 0.0171 & 0.0117 & 0.0542 & 0.0372 & 0.2906 & 0.2067 & 0.2146 & 0.1439 & 0.3217 & 0.2221  \\
VIIDEO~\cite{mittal2015completely} & 0.0141 & 0.0075 & 0.0918 & 0.0756 & 0.0124 & 0.0126 & 0.0484 & 0.0300 & 0.0349 & 0.0178 & 0.0776 & 0.0547 & 0.2104 & 0.1389  \\
ChipQA~\cite{ebenezer2021chipqa} & 0.2418 & 0.1749 & 0.1901 & 0.1342 & 0.3087 & 0.2253 & 0.2894 & 0.2071 & 0.2112 & 0.1532 & 0.2099 & 0.1534 & 0.3780 & 0.2708  \\
NR-CUGCVQA~\cite{li2021full} & \underline{0.5063} & \underline{0.3463} & 0.0989 & 0.0638 & 0.3045 & 0.1978 & \underline{0.5790} & \underline{0.4068} & \underline{0.6470} & \underline{0.4567} & \textbf{0.6467} & \textbf{0.4613} & \underline{0.4927} & \underline{0.3450}  \\
NR-CONTRIQUE~\cite{madhusudana2022image} & 0.3268 & 0.2208 & \textbf{0.2645} & \textbf{0.1797} & 0.3390 & 0.2315 & 0.3827 & 0.2582 & 0.4485 & 0.3038 & 0.3458 & 0.2299 & 0.3890 & 0.2617  \\
SimpleVQA~\cite{sun2022deep} & \textbf{0.5390} & \textbf{0.3741} & 0.1512 & 0.1043 & \textbf{0.5218} & \textbf{0.3595} & \textbf{0.5862} & \textbf{0.4200} & 0.5877 & 0.4132 & 0.5494 & 0.3839 & \textbf{0.6097} & \textbf{0.4221}  \\
FastVQA~\cite{wu2022fast} & 0.0232 & 0.0163 & 0.0443 & 0.0295 & 0.0465 & 0.0301 & 0.0060 & 0.0023 & 0.0065 & 0.0045 & 0.0442 & 0.0316 & 0.0067 & 0.0038  \\
FasterVQA~\cite{wu2023neighbourhood} & 0.0804 & 0.0535 & 0.0160 & 0.0097 & 0.0741 & 0.0487 & 0.0110 & 0.0060 & 0.0856 & 0.0573 & 0.0752 & 0.0500 & 0.1026 & 0.0685  \\
CONVIQT~\cite{madhusudana2023conviqt} & 0.4934 & 0.3374 & 0.1630 & 0.1107 & \underline{0.3760} & \underline{0.2530} & 0.4857 & 0.3304 & \textbf{0.7132} & \textbf{0.5266} & \underline{0.6028} & \underline{0.4145} & 0.3735 & 0.2525  \\
\bottomrule
\end{tabular}
 }
 \label{tab:frresults}
\vspace{-5pt}
\end{table*}

\begin{table*}[t]
\caption{The correlation results with MOS of the tested no-reference metrics on BVI-UGC across different reference and codec groups. In each group, the best and the second best metrics are \textbf{boldfaced} and \underline{underlined}.}
\centering
\resizebox{\linewidth}{!}{
\begin{tabular}{r||l|l||l|l||l|l||l|l||l|l||l|l||l|l}
\toprule
\multirow{2}{*}{\textbf{Databases}} &  \multicolumn{2}{c||}{\textbf{BVI-UGC}} &  \multicolumn{2}{c||}{\textbf{BVI-UGC \textcolor{red!85!black}{low}}} &  \multicolumn{2}{c||}{\textbf{BVI-UGC \textcolor{myyellow}{medium}}} &  \multicolumn{2}{c||}{\textbf{BVI-UGC \textcolor{green!70!black}{high}}} &  \multicolumn{2}{c||}{\textbf{BVI-UGC H264}} &  \multicolumn{2}{c||}{\textbf{BVI-UGC H265}} &  \multicolumn{2}{c}{\textbf{BVI-UGC libaom}} \\
\cmidrule{2-15} 
& \textbf{SROCC} & \textbf{KRCC} & \textbf{SROCC} & \textbf{KRCC} & \textbf{SROCC} & \textbf{KRCC} & \textbf{SROCC} & \textbf{KRCC} & \textbf{SROCC} & \textbf{KRCC} & \textbf{SROCC} & \textbf{KRCC} & \textbf{SROCC} & \textbf{KRCC} \\
\midrule \multicolumn{5}{l}{\textbf{No-reference} (measuring absolute quality, to correlate with MOS)} \\ 
\midrule
NIQE~\cite{mittal2012making} & 0.2533 & 0.1700 & \textbf{0.3816} & \textbf{0.2558} & 0.2794 & 0.1871 & 0.1117 & 0.0737 & 0.4077 & 0.2837 & 0.3184 & 0.2155 & 0.2266 & 0.1532  \\
BRISQUE~\cite{mittal2012no} & 0.2527 & 0.1758 & 0.2313 & 0.1465 & 0.3215 & 0.2200 & 0.3333 & 0.2318 & 0.2498 & 0.1761 & 0.2356 & 0.1667 & 0.2860 & 0.1991  \\
VBLIINDS~\cite{saad2014blind} & 0.1211 & 0.0808 & 0.0565 & 0.0361 & 0.1355 & 0.0911 & 0.0143 & 0.0106 & 0.1384 & 0.0956 & 0.1516 & 0.1017 & 0.3513 & 0.2370  \\
VIIDEO~\cite{mittal2015completely} & 0.1651 & 0.1102 & 0.2878 & 0.1940 & 0.1322 & 0.0881 & 0.0261 & 0.0181 & 0.1565 & 0.1055 & 0.2904 & 0.1943 & 0.1450 & 0.0951  \\
ChipQA~\cite{ebenezer2021chipqa} & 0.1994 & 0.1464 & 0.0652 & 0.0419 & 0.2687 & 0.1951 & 0.3143 & 0.2313 & 0.1754 & 0.1292 & 0.1642 & 0.1212 & 0.3040 & 0.2245  \\
NR-CUGCVQA~\cite{li2021full} & 0.4333 & 0.2953 & 0.0664 & 0.0469 & 0.3649 & 0.2479 & \underline{0.4723} & \underline{0.3191} & \textbf{0.6202} & \textbf{0.4433} & \underline{0.5548} & \underline{0.3904} & 0.2940 & 0.1977  \\
NR-CONTRIQUE~\cite{madhusudana2022image} & 0.3376 & 0.2294 & 0.2076 & 0.1417 & 0.3738 & 0.2552 & 0.3056 & 0.2113 & 0.3023 & 0.2045 & 0.3514 & 0.2420 & \underline{0.4321} & \underline{0.2977}  \\
SimpleVQA~\cite{sun2022deep} & \textbf{0.5203} & \textbf{0.3641} & 0.1439 & 0.0983 & \textbf{0.5562} & \textbf{0.3933} & \textbf{0.5345} & \textbf{0.3702} & \underline{0.5832} & \underline{0.4186} & 0.5391 & 0.3821 & \textbf{0.5586} & \textbf{0.3926}  \\
FastVQA~\cite{wu2022fast} & 0.1897 & 0.1266 & 0.3417 & 0.2236 & 0.2905 & 0.1951 & 0.1539 & 0.1033 & 0.2186 & 0.1455 & 0.2091 & 0.1405 & 0.1974 & 0.1333  \\
FasterVQA~\cite{wu2023neighbourhood} & 0.1922 & 0.1285 & 0.3215 & 0.2051 & 0.2933 & 0.1966 & 0.1470 & 0.0984 & 0.2223 & 0.1483 & 0.2116 & 0.1422 & 0.1889 & 0.1285  \\
CONVIQT~\cite{madhusudana2023conviqt} & \underline{0.4678} & \underline{0.3232} & \underline{0.3507} & \underline{0.2323} & \underline{0.4855} & \underline{0.3359} & 0.3594 & 0.2469 & 0.5387 & 0.3773 & \textbf{0.5590} & \textbf{0.3927} & 0.4002 & 0.2750  \\
\bottomrule
\end{tabular}
 }
 \label{tab:nrresults}
\vspace{-5pt}
\end{table*}

\begin{figure*}[htb]
\small
\centering
\begin{minipage}[b]{0.32\linewidth}
  \centering
  \centerline{\includegraphics[width=\textwidth]{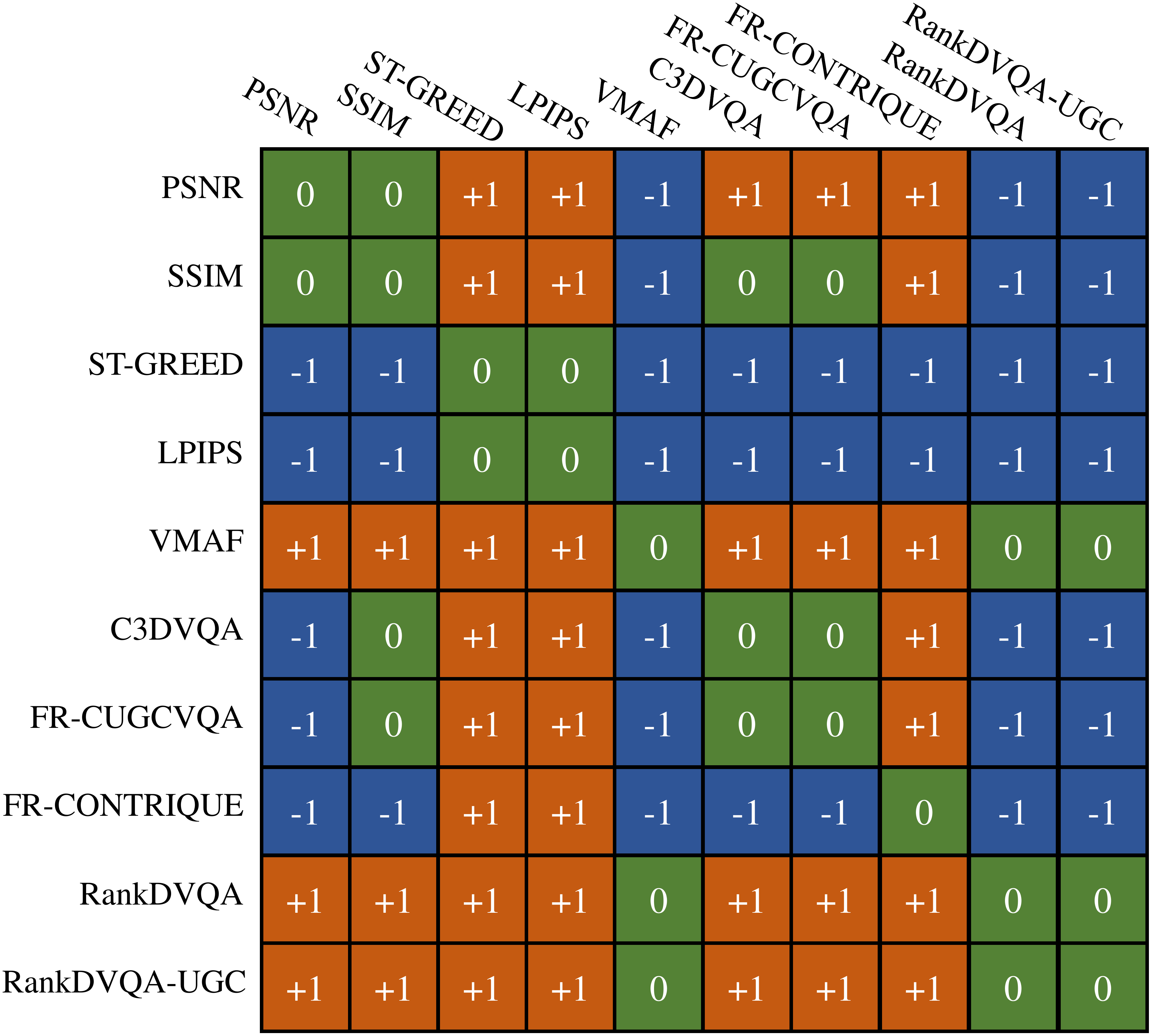}}
(a) FR metrics
\end{minipage}
\begin{minipage}[b]{0.32\linewidth}
  \centering
  \centerline{\includegraphics[width=\textwidth]{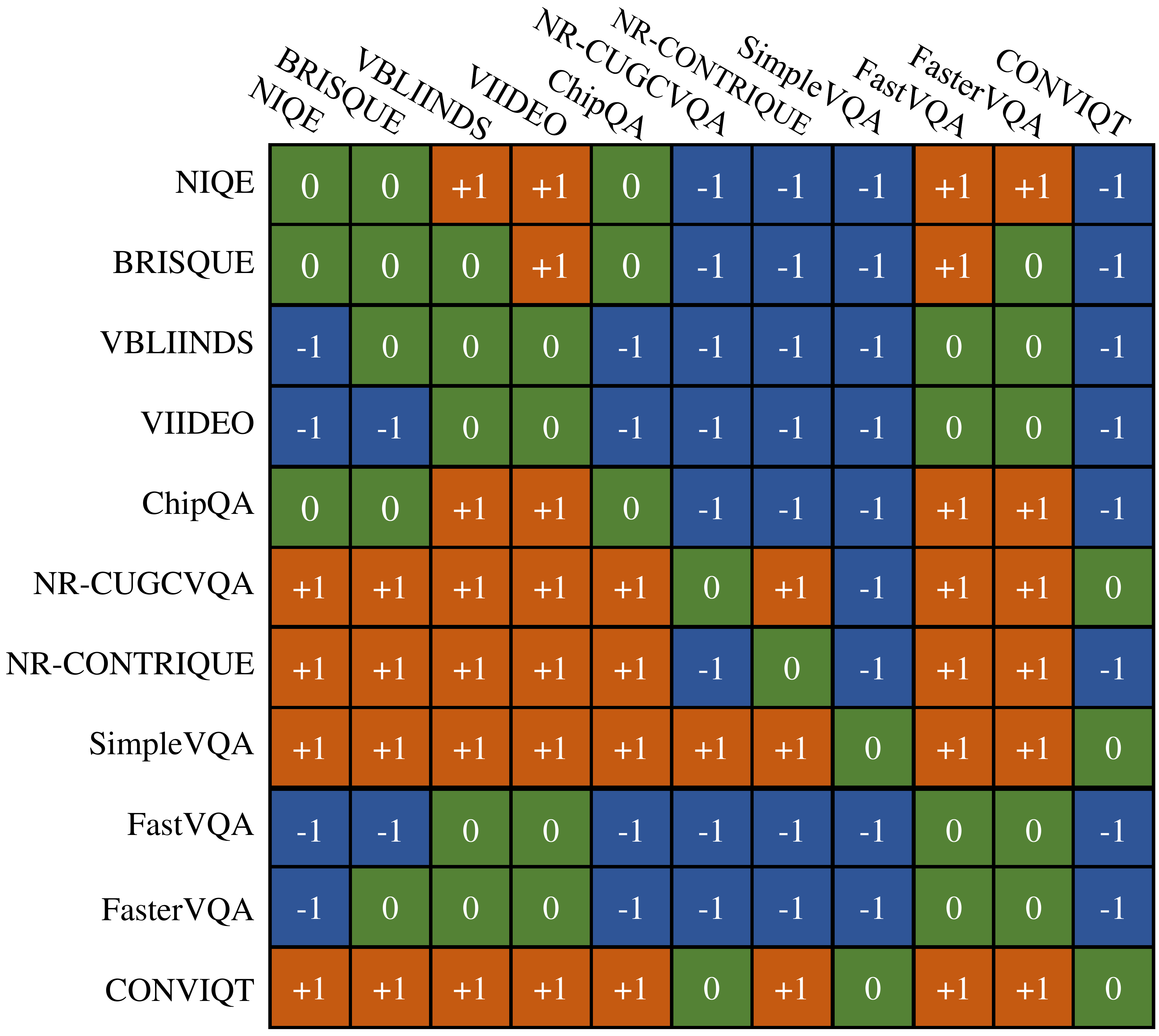}}
(b) NR metrics (against DMOS)
\end{minipage}
\begin{minipage}[b]{0.32\linewidth}
  \centering
  \centerline{\includegraphics[width=\textwidth]{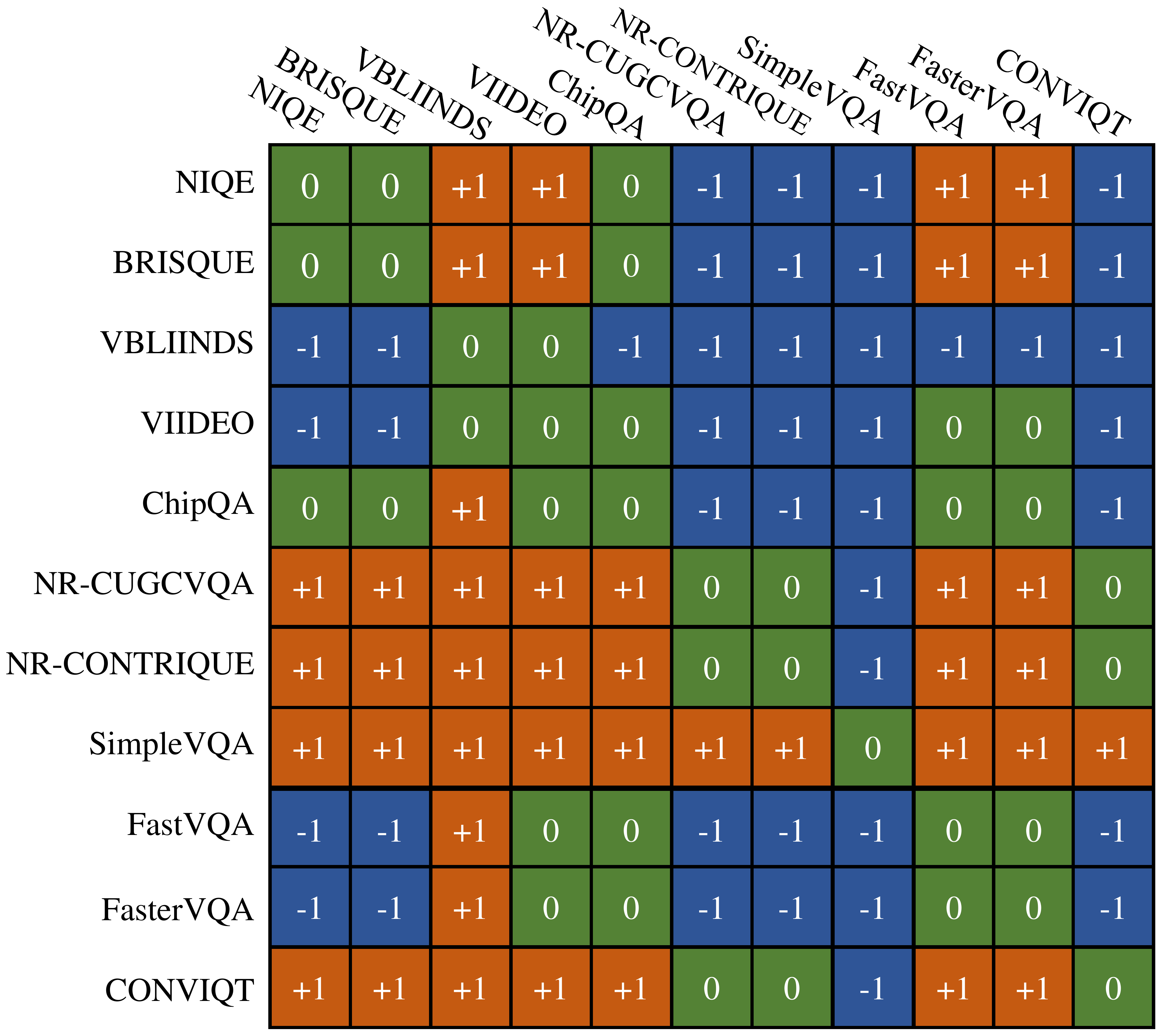}}
(c) NR metrics (against MOS)
\end{minipage}
\caption{Pairwise comparisons between the overall performances. The color and value of the cells indicate the F-test results between the DMOS prediction residuals of the metrics pair, at a 95\% confidence interval. Orange cell with +1 value indicates the metric in the row is superior to the metric in the column and blue cell with -1 value means the opposite. Green cell with 0 value denotes statistical equivalence.}
\label{fig_ftest}
\end{figure*}

\begin{figure*}[!t]
\hfill
\centering
  \small
\begin{minipage}{0.24\linewidth}
  \centering
  \centerline{\includegraphics[width=1.05\linewidth]{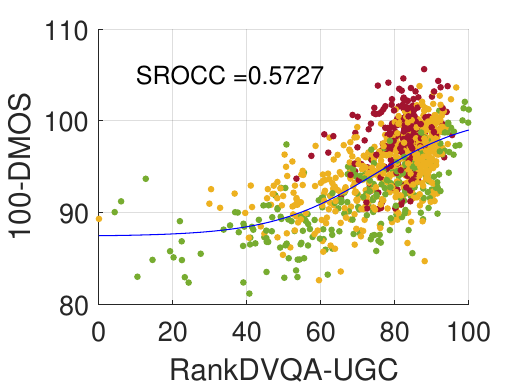}}
  (a)
\end{minipage}
\begin{minipage}{0.24\linewidth}
  \centering
  \centerline{\includegraphics[width=1.05\linewidth]{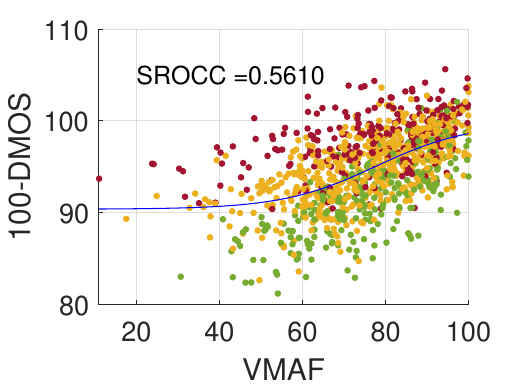}}
  (b)
\end{minipage}
\begin{minipage}{0.24\linewidth}
  \centering
  \centerline{\includegraphics[width=1.05\linewidth]{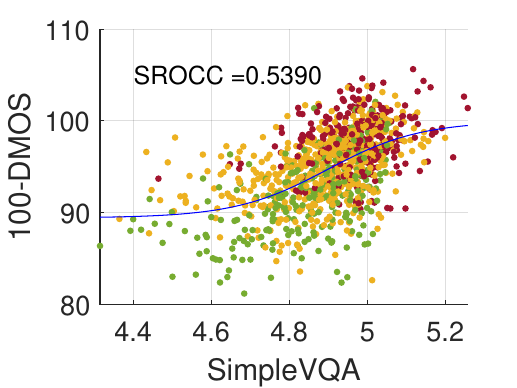}}
  (c)
\end{minipage}
\begin{minipage}{0.24\linewidth}
  \centering
  \centerline{\includegraphics[width=1.05\linewidth]{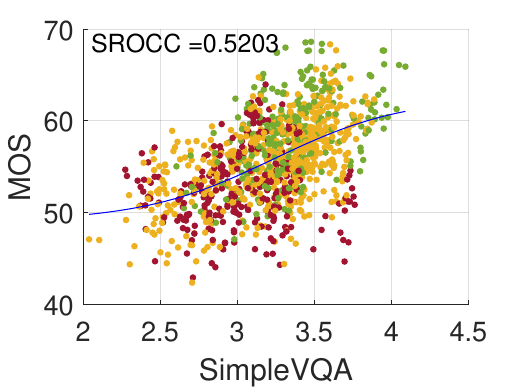}}
  (d)
\end{minipage}
\caption{Scatter plots of subjective scores against the predictions of the best-performing models. The \textcolor{green!70!black}{\textbf{green}}, \textcolor{myyellow}{\textbf{yellow}}, and \textcolor{red!85!black}{\textbf{red}} scatter points correspond to the transcoded sequences generated from three groups with different reference qualities, high, medium, and low. The blue lines are the logistic functions fitted between the model predictions and subjective scores on the entire BVI-UGC database.} 
\vspace{-10pt}
\label{fig_fittings}
\end{figure*}

The BVI-UGC database with its associated subjective data contributes a rigorous and valuable benchmarking tool for evaluating the performance of quality assessment methods on user-generated content. More importantly, due to its unique design, it can be used to test both full-reference and no-reference metrics in the context of transcoding. This is achieved by calculating the correlation between quality indices generated by the objective quality models and the DMOS of transcoded content. It can also be employed,  like many other UGC databases such as YouTube UGC~\cite{wang2019youtube}, KoNViD-1k~\cite{konvid1k,hosu2017konstanz}, KonViD-150k~\cite{konvid150k,hahn2021}, and LIVE-VQC~\cite{sinno2018large}, to assess the performance of no-reference metrics directly (against MOS) without references, simulating the quality of experience estimation at the user end. 

\subsection{The benchmarked VQA methods}

In this section, we present a comprehensive evaluation with 10 popular full-reference (FR) and 11 no-reference (NR) metrics. Specifically, the tested FR VQA methods include PSNR, SSIM~\cite{wang2004image}, MS-SSIM~\cite{wang2003multiscale}, ST-GREED~\cite{madhusudana2021st}, LPIPS~\cite{zhang2018unreasonable}, VMAF~\cite{li2016toward}, C3DVQA~\cite{xu2020c3dvqa}, FR-CUGCVQA~\cite{li2021full}, FR-CONTRIQUE~\cite{madhusudana2022image},  RankDVQA~\cite{feng2024rankdvqa} and RankDVQA-UGC~\cite{qi2023full}. Among these methods, PSNR, SSIM, MS-SSIM are conventional quality metrics, VMAF and ST-GREED are regression-based models, and LPIPS, CONTRIQUE, CUGCVQA, C3DVQA and RankDVQA are deep video quality assessment. It is noted that, among the metrics above, CUGCVQA and CONTRIQUE have both full-reference and no-reference implementations. All FR quality metrics tested here are used to calculate the quality differences between the transcoded sequences and their corresponding non-pristine reference sequences. The Spearman Ranking Order Correlation Coefficients (SROCC) and Kendall Ranking Correlation Coefficients (KRCC) between the predicted quality difference values and their corresponding DMOS are then computed to measure their performance.

Similarly, 11 NR quality metrics also include conventional or machine learning-based methods such as NIQE~\cite{mittal2012making}, BRISQUE~\cite{mittal2012no}, VBLIINDS~\cite{saad2014blind}, VIIDEO~\cite{mittal2015completely}, and deep learning-based quality models, e.g., ChipQA~\cite{ebenezer2021chipqa}, NR-CUGCVQA~\cite{li2021full}, NR-CONTRIQUE~\cite{madhusudana2022image}, SimpleVQA~\cite{sun2022deep}, CONVIQT~\cite{madhusudana2023conviqt}, FastVQA~\cite{wu2022fast} and FasterVQA~\cite{wu2023neighbourhood}. For each NR metric, we first adapt to the FR transcoding scenario by calculating the quality indices of transcoded videos and its non-pristine reference separately, and obtaining their quality differences. This is to measure the quality degradation of the transcoded content relative to the reference. We also assess the performance of these NR models by calculating their correlation coefficients between their predicted quality indices and the corresponding MOS of the transcoded sequences. 

For all the learning-based methods, including LPIPS, VMAF, ChipQA, CUGCVQA, CONTRIQUE, ST-GREED, C3DVQA, RankDVQA, BRISQUE, SimpleVQA, CONVIQT, FastVQA and FasterVQA, their pre-trained models provided in the associated original literature were used in this experiment. To test model generalization, we did not perform any cross-validation within the proposed database.

\subsection{Evaluation results}

\autoref{tab:frresults} summarizes the performance results of all the tested full-/no-reference VQA methods in the context of transcoding, where the DMOS values are employed to calculate the correlation coefficients. It is noted that none of the tested quality assessment methods achieve satisfactory overall correlation performance on BVI-UGC - the best performer RankDVQA-UGC only offers a SROCC value of 0.5727. Among all the no-reference quality metrics, SimpleVQA achieves the highest SROCC value both when measuring the quality degradation (0.5390) and when predicting the absolute quality (0.5203). We have further divided the whole BVI-UGC database into three subsets according to the unpristing references. It can be observed that for test sequences generated from low-quality references, most full-reference quality models perform worse than on those from medium-/high-quality references. This confirms our assumption that the quality of reference videos does affect the quality prediction accuracy of full-reference quality models. We performed another segmentation of the database based on the codec used in transcoding, and found that many quality metrics achieve better performance on H.264 or H.265 content compared to libaom compressed sequences. This may be because of the fact that these VQA methods are trained and/or validated more often on H.264/H.265 compressed content. 

As mentioned above, the BVI-UGC database can also be used to evaluate no-reference quality metrics in terms of their ability to directly predict the visual quality of distorted videos (using MOS to calculate correlation coefficients). \autoref{tab:nrresults} summarizes their results when benchmarked on the BVI-UGC database. Here the best overall performance is also provided by SimpleVQA, while the second best performer is CONVIQT. Based on all these results we can conclude that assessing the perceptual quality of UGC content is a highly challenging task, in particular when the reference content is distorted. More advanced VQA models are urgently required to offer enhanced prediction performance.

To further validate the performance ranking among the benchmarked quality assessment methods, we also conducted an F-test between every two metrics to check the statistical significance of their difference, following the practice in ~\cite{seshadrinathan2009motion,seshadrinathan2010study}. Specifically, a pairwise comparison was performed on the residuals between the DMOS (or MOS) and the model prediction after non-linear regression~\cite{video2000final}. \autoref{fig_ftest} (a-c) summarize the F-test results between full-/no-reference metrics (on DMOS) and no-reference models (on MOS). We can observe that RankDVQA-UGC, RankDVQA and VMAF are the best performers among full-reference metrics, all of which significantly outperform seven other quality models in a 95\% confidence interval. For no-reference quality models, SimpleVQA is statistically better than ten other NR VQA methods based on DMOS or MOS. 

In order to analyze the correlation performance of the objective quality metrics, the scatter plots of the predictions of the selected, well-performing models against subjective scores, along with the fitted logistic curves, are shown in ~\autoref{fig_fittings}. It can be observed that, in all cases, the scatter points are distributed sparsely along the fitting curves, which also demonstrates the unsatisfactory performance of existing VQA methods on this database from a different perspective. 

All the results shown in this section confirm the urgent need for an accurate and robust quality metric, which can adapt to various reference content scenarios and different distortion types, to facilitate UGC streaming applications.

\section{Conclusion}
\label{sec:conclusion}

In this paper, we have presented a novel video database, BVI-UGC, which is the first UGC database to contain (non-pristine) references with various levels of distortions and transcoded content generated by multiple codecs. It consists of 60 pseudo-pristine source sequences with diverse and representative user-generated video content, covering 15 popular UGC categories. These were further used to produce 60 non-pristine reference sequences and 1,080 transcoded sequences following a typical UGC streaming pipeline. Based on this database, we designed and performed a large-scale crowdsourcing subjective study on the perceptual quality of of both non-pristine referece and transcoded videos. The collected subjective scores, together with the video clips, have been employed to benchmark the performance of 21 full-reference and no-reference popular quality assessment methods. The results clearly show that all these quality metrics fail to perform well on this database, with ranking order correlation coefficient (SROCC) values below 0.6. 

We believe that the BVI-UGC database will provide a valuable resource to the research community for developing and validating new video quality assessment models in the context of UGC transcoding. Future work is now required to investigate full-reference and no-reference quality metrics, which can predict the perceived quality of streamed UGC content more accurately and robustly.

\small
\bibliographystyle{IEEEtran}
\bibliography{BVI-UGC_TIP}

\end{document}